\documentclass[aps,prd,twocolumn,groupedaddress,nofootinbib,longbibliography]{revtex4-1}

\usepackage{amssymb}
\usepackage{amsmath}
\usepackage[linktocpage=true]{hyperref}
\usepackage{graphicx}

\newcommand{\eqH}{\stackrel{\rm H}{=}}

\begin{document}

\title{On geometry of deformed black holes:\\
       I. Majumdar-Papapetrou binary}

\author{O. Semer\'ak}
\email[]{oldrich.semerak@mff.cuni.cz}

\author{M. Basovn\'{\i}k}
\email[]{mbasovnik@gmail.com}

\affiliation{Institute of Theoretical Physics, Faculty of Mathematics and Physics,
             Charles University in Prague, Czech Republic}

\date{\today}

\begin{abstract}
Although black holes are eminent manifestations of very strong gravity, the geometry of space-time around and even inside them can be significantly affected by additional bodies present in their surroundings. We study such an influence within static and axially symmetric (electro-)vacuum space-times described by exact solutions of Einstein's equations, considering astrophysically motivated configurations (such as black holes surrounded by rings) as well as those of pure academic interest (such as specifically ``tuned" systems of multiple black holes). The geometry is represented by the simplest invariants determined by the metric (the lapse function) and its gradient (gravitational acceleration), with special emphasis given to curvature (the Kretschmann and Ricci-square scalars). These quantities are analyzed and their level surfaces plotted both above and below the black-hole horizons, in particular near the central singularities.
Estimating that the black hole could be most strongly affected by the other black hole, we focus, in this first paper,
on the Majumdar--Papapetrou solution for a binary black hole and compare the deformation caused by ``the other" hole (and the electrostatic field) with that induced by rotational dragging in the well-known Kerr and Kerr--Newman solutions.
\end{abstract}

\pacs{0420Jb, 0440Nr, 0470Bw}

\maketitle

\section{Introduction}

``Although it appears that the most exciting future development in black hole theory will be concerned with dynamic aspects, there remains a great deal to be done in stationary black hole theory, particularly in relation with non-vacuum black holes." B. Carter did not mention quantum aspects in his 1972 lecture \cite{Carter-73}, but otherwise the sentence remains valid. However, dynamical black-hole processes do not give much opportunities to exact analytical solution, they are rather being tackled by numerical and approximation methods, and even in stationary but non-vacuum cases the compass of detailed exact analytical treatment is restricted, namely to axially symmetric (and ideally also static) configurations.

Black holes are the most conservative, today almost routine explanation of a whole bunch of high-energy astrophysical phenomena. Yet even solitary holes remain hard to imagine (though Chandrasekhar \cite{Chandrasekhar} considered them the simplest objects in the universe), in particular, if they are spinning so fast as often supposed in galactic nuclei and in some X-ray binaries, their horizon, taken as a 2D surface at any fixed Killing time, is partially a surface of negative curvature. Moreover, the ``observed" holes must be strongly interacting with matter and electromagnetic fields. In the astrophysical models the gravitational effect of these is neglected, thus space-time is assumed to have a Kerr form corresponding to an {\em isolated} rotating black hole. It is indeed likely that the accreting material is too light to have any significant effect on the gravitational potential, but it may well contribute to higher derivatives of the field, the more so if it is collapsed into a thin disc or even a ring (see e.g. \cite{Semerak-04} and references therein). Hence, the space-time curvature around and probably even inside the black hole might be modified significantly by the ambient matter.

This conjecture has been confirmed by \cite{FrolovS-07} on a Schwarzschild black hole ``subject to" higher gravitational multipoles. (The authors also extended the results to a charged black hole in \cite{AbdolrahimiFS-09}.) In particular, it is known that the central singularity of {\em static} black holes remains spatially point-like and that the structure of the whole space-time remains similar, irrespectively of the external influence \cite{GerochH-82}, but the paper \cite{FrolovS-07} pointed out that the vicinity of the singularity may still be deformed considerably. Actually, it was showed there that the central region of strongest curvature can be stretched in such an (anisotropic) way that it may even reach above the horizon in certain (though rather extreme) circumstances.

In the present work, we study the effect of the additional source on the black-hole geometry by calculating and plotting several invariants determined by the metric (the lapse function and the azimuthal-circumference radius), by its gradient (an analogue of the Newtonian gravitational acceleration, known as the surface gravity when evaluated on the horizon) and by the Riemann tensor (Kretschmann scalar and similar quadratic scalar obtained from the Ricci tensor); the quantities are reminded in section \ref{scalars}. We choose two static and axially symmetric exact space-times, the Majumdar--Papapetrou electro-vacuum solution with just two black holes (section \ref{MP}) and the vacuum solution given by ``superposition" of a Schwarzschild black hole surrounded by a concentric Bach--Weyl thin ring (next paper). These two situations seem to promise a noticeable influence on the black hole, since the additional sources considered -- another black hole in the Majumdar--Papapetrou solution and thin (two-dimensional) ring in the second case -- are one of the strongest possible gravitational sources. In order to compare the distortion caused by these external sources with the distortion induced by rotation, we however first repeat some geometrical properties of the Kerr(--Newman) black holes in section \ref{Kerr}. Concluding remarks are added in section \ref{concluding}.

We use geometrized units in which $c=1$, $G=1$, index-posed comma/semicolon indicates partial/covariant derivative and usual summation rule is employed. Signature of the space-time metric $g_{\mu\nu}$ is ($-$+++), Riemann tensor is defined according to $V_{\nu;\kappa\lambda}-V_{\nu;\lambda\kappa}={R^\mu}_{\nu\kappa\lambda}V_\mu$
and Ricci tensor by $R_{\nu\lambda}={R^\kappa}_{\nu\kappa\lambda}$.
The equations which are only valid on a black-hole horizon are written with the index `H', $X\eqH Y$.

Let us note, finally, that the real astrophysical black holes are also different from their Kerr ideals due to the whole non-vacuum universe around, not only due to the nearby accreting matter and fields, of course. We however do not take the cosmological setting into account and, in particular, we set the cosmological constant at zero.

\section{Summary on scalars considered}
\label{scalars}

Every stationary, axially symmetric and orthogonally transitive space-time can be described by the metric
\[{\rm d}s^2=-N^2{\rm d}t^2+g_{\phi\phi}({\rm d}\phi-\omega{\rm d}t)^2
             +g_{11}({\rm d}x^1)^2+g_{22}({\rm d}x^2)^2 \,,\]
where the coordinates $t$ and $\phi$ are adapted to space-time symmetries
(thus $N$, $g_{\phi\phi}$, $\omega$ and $g_{11}$, $g_{22}$ do not depend on them),
$\eta^\mu\equiv\partial x^\mu/\partial t$ and $\xi^\mu\equiv\partial x^\mu/\partial\phi$
being the time and azimuthal Killing vector fields, the coordinates $x^1$ and $x^2$ cover the meridional planes, orthogonal to both Killing directions (and existing, as integral surfaces, due to the orthogonal transitivity), the lapse function $N$ is determined by $N^2\equiv -g_{tt}-g_{t\phi}\omega$ and the function $\omega$ is given by $\omega\equiv -g_{t\phi}/g_{\phi\phi}$.
The simplest invariants of the above metric, the inner products of the Killing vectors
\[g_{\mu\nu}\eta^\mu\eta^\nu=g_{tt} \,, \quad
  g_{\mu\nu}\xi^\mu\xi^\nu=g_{\phi\phi} \,, \quad
  g_{\mu\nu}\eta^\mu\xi^\nu=g_{t\phi} \,,\]
are usually being represented in terms of their more intuitive combinations -- the lapse $N$ (dilation factor of the zero-angular-momentum observer), the dragging potential $\omega$ (representing angular velocity of rotational frame dragging) and the azimuthal-circumference radius $\sqrt{g_{\phi\phi}}\,$.
In a {\em static} case, there is no dragging, $\omega=0$, so $N^2=-g_{tt}$.

The most useful simple scalar given by gradient of the metric is
\begin{equation}  \label{kappa2}
  \kappa^2\equiv g^{\mu\nu}N_{,\mu}N_{,\nu}
          =g^{11}(N_{,1})^2+g^{22}(N_{,2})^2 \,.
\end{equation}
$\kappa$ is an analogue of the magnitude of Newtonian gravitational acceleration and on the black-hole horizon it is known as surface gravity (on {\em stationary} horizons it is uniform, which is the case here).

On the level of curvature (second derivatives of the metric), one can find 14 algebraically independent invariants. In the vacuum case, only two quadratic and two cubic invariants are left,
\begin{align*}
  &R_{\mu\nu\kappa\lambda}R^{\mu\nu\kappa\lambda}\equiv K \;\;\;\;
     {\rm \dots~Kretschmann~scalar}\;, \\
  &{^*\!R}_{\mu\nu\kappa\lambda}R^{\mu\nu\kappa\lambda}\equiv {^*\!K} \;\;\;\;
     \mbox{{\rm \dots~Chern--Pontryagin~scalar}} \;, \\
  &{R^{\mu\nu}}_{\kappa\lambda}{R^{\kappa\lambda}}_{\alpha\beta}
   {R^{\alpha\beta}}_{\mu\nu} \;, \;\;\;\;
   {{^*\!R}^{\mu\nu}}_{\kappa\lambda}
   {R^{\kappa\lambda}}_{\alpha\beta}{R^{\alpha\beta}}_{\mu\nu} \;,
\end{align*}
where ${R^\mu}_{\nu\kappa\lambda}$ is the Riemann tensor and
${^*\!R}_{\mu\nu\kappa\lambda}\equiv
 \frac{1}{2}\,\epsilon_{\mu\nu\alpha\beta}{R^{\alpha\beta}}_{\kappa\lambda}$
is its left dual.
In the non-vacuum case, the remaining 10 scalars are determined by the Ricci tensor.
For a {\em static} space-time, the scalars given by Riemann-tensor dual vanish.

Also worth recalling is the special case with just source-free electromagnetic field present (called electro-vacuum case) when the energy-momentum tensor reads
\begin{equation}  \label{Tmunu,EM}
  T_{\mu\nu}=\frac{1}{4\pi}\left(F_{\mu\lambda}{F_\nu}^\lambda-
                                 \frac{1}{4}\,g_{\mu\nu}F_{\kappa\lambda}F^{\kappa\lambda}\right),
\end{equation}
with $F_{\mu\nu}\equiv A_{\nu,\mu}-A_{\mu,\nu}$ denoting the electromagnetic-field tensor and $A_\mu$ the electromagnetic four-potential.
Such an energy-momentum tensor is traceless, $T^\nu_\nu=0$, so the Einstein equations (without the cosmological term) imply that the Ricci scalar $R\equiv R^\nu_\nu$ is zero, too. It is also well known that $F_{\mu\nu}$ yields just two non-trivial and independent invariants, $F_{\mu\nu}F^{\mu\nu}$ and $F_{\mu\nu}{^*\!F}^{\mu\nu}$, of which the second, given by the dual tensor ${^*\!F}^{\mu\nu}\equiv\frac{1}{2}\,\epsilon^{\mu\nu\alpha\beta}F_{\alpha\beta}$,
vanishes in a static situation.

Hence, in a {\em non-static} (but stationary) {\em vacuum} space-times, there are only two quadratic curvature invariants, the Kretschmann scalar and the Chern--Pontryagin scalar.
On the other hand, in a {\em static electro-vacuum} space-times, there are again only two quadratic curvature invariants, the Kretschmann scalar and the trace of the Ricci-tensor square
\begin{align}
  R_{\mu\nu}R^{\mu\nu}
  &= 4\,F_{\mu\lambda}F^{\nu\lambda}F_{\nu\kappa}F^{\mu\kappa}-(F_{\mu\nu}F^{\mu\nu})^2 = \nonumber\\
  &= (F_{\mu\nu}F^{\mu\nu})^2+(F_{\mu\nu}{^*\!F}^{\mu\nu})^2
   = (F_{\mu\nu}F^{\mu\nu})^2 \,.
\end{align}

\subsection{Basic scalars in a static axisymmetric electro-vacuum}

We add several remarks to simplifications and problems that occur in computation of the above invariants in static (and axially symmetric) electro-vacuum space-times.
For an extreme horizon, there is no dynamical region, so the lapse squared $N^2=-g_{tt}$ as well as $\kappa^2=g^{ii}(N_{,i})^2$ are nowhere negative.
Below a non-extreme horizon, $N^2=-g_{tt}$ is {\em negative}, so $N^2$ itself or $|N|=\sqrt{g_{tt}}$ has to be treated there instead of $N=\sqrt{-g_{tt}}$. Consequently, since $N$ is pure imaginary there, its gradient is also pure imaginary, so for our diagonal metric $\kappa^2=g^{11}(N_{,1})^2+g^{22}(N_{,2})^2$ is real everywhere. One can imagine now that we use as $x^1$ some radial coordinate which is constant all over the horizon, and as $x^2$ the usual latitudinal coordinate $\theta$ which ranges from $\theta=0$ to $\theta=\pi$ which correspond to the opposite halves of the symmetry axis. Then $N_{,\theta}=0$ at $\theta=0$ and if the space-time is reflection symmetric with respect to the $\theta=\pi/2$ plane (called equatorial), $N_{,\theta}=0$ at $\theta=\pi/2$ as well (the latter would not hold if there was some mass-line{\/}/{\/}mass-shell along these locations, but there cannot be any below a regular static horizon). Hence, at $\theta=0$ (and $\theta=\pi/2$) one has just $\kappa^2=g^{rr}(N_{,r})^2$, which is {\em positive} because $g^{rr}<0$ below horizon. This indicates that $\kappa$ remains real even inside a non-extreme black hole.

Let us repeat basic facts on Weyl solutions \cite{Weyl-17} now (see Appendix C of \cite{BiniGR-07}).
If a space-time is static and axially symmetric, then in regions where the energy-momentum tensor satisfies $T^1_1+T^2_2=0$ (remind that $x^1$, $x^2$ cover the meridional planes orthogonal to Killing directions $t$, $\phi$) the metric can be written in the Weyl form
\begin{equation}  \label{Weyl-metric}
  {\rm d}s^2=-e^{2\nu}{\rm d}t^2+\rho^2 e^{-2\nu}{\rm d}\phi^2
             +e^{2\lambda-2\nu}({\rm d}\rho^2+{\rm d}z^2) \,,
\end{equation}
where the unknown functions $\nu$ and $\lambda$ only depend on cylindrical-type radius $x^1\equiv\rho$ and the ``vertical" linear coordinate $x^2\equiv z$ which cover the meridional planes in an isotropic manner. The simplest metric scalars are obvious, $N=e^{\nu}$ and $\sqrt{g_{\phi\phi}}=\rho/N$.
Einstein's field equations reduce to (e.g. \cite{Carminati-81})
\begin{align}
  \nu_{,\rho\rho}+\frac{\nu_{,\rho}}{\rho}+\nu_{,zz}
    &= 4\pi e^{2\lambda-2\nu}(T^\phi_\phi-T^t_t) \\
    &= e^{2\lambda-2\nu}(F_{\phi\lambda}F^{\phi\lambda}-F_{t\lambda}F^{t\lambda}) \\
    &= e^{-2\nu}\left[(\Phi_{,\rho})^2+(\Phi_{,z})^2\right], \label{Laplace,nu,elstat}
\end{align}
\begin{align}
  \lambda_{,\rho}-\rho(\nu_{,\rho})^2+\rho(&\nu_{,z})^2
     = 4\pi\rho\,(T_{\rho\rho}-T_{zz}) \\
    &= \rho\,(F_{\rho\lambda}{F_\rho}^\lambda-F_{z\lambda}{F_z}^\lambda) \\
    &= {-}\rho e^{-2\nu}\left[(\Phi_{,\rho})^2-(\Phi_{,z})^2\right], \label{Weyl-lambda,rho,elstat} \\
  \lambda_{,z}-2\rho\nu_{,\rho}\nu_{,z}
    &= 8\pi\rho\,T_{\rho z} \\
    &= 2\rho\,F_{\rho\lambda}{F_z}^\lambda \\
    &= {-}2\rho e^{-2\nu}\Phi_{,\rho}\Phi_{,z} \,, \label{Weyl-lambda,z,elstat}
\end{align}
\begin{align}
  \lambda_{,\rho\rho}+\lambda_{,zz}+(&\nu_{,\rho})^2+(\nu_{,z})^2
     = 8\pi e^{2\lambda-2\nu}T^\phi_\phi \\
    &= \frac{1}{2}\,e^{2\lambda-2\nu}(4F_{\phi\lambda}F^{\phi\lambda}-F_{\mu\nu}F^{\mu\nu}) \\
    &= e^{-2\nu}\left[(\Phi_{,\rho})^2+(\Phi_{,z})^2\right], \label{Weyl,eq4,elstat}
\end{align}
where the second forms of the r.h. sides specialize to the pure-electromagnetic energy-momentum tensor (\ref{Tmunu,EM}) and the third forms are obtained after restriction to the electrostatic situation when the electromagnetic field can be expressed in terms of a scalar potential $\Phi(\rho,z)$ as
\begin{equation}
  A_\mu=(-\Phi,0,0,0)
  \quad \Longrightarrow \;\;
  F_{t\rho}=\Phi_{,\rho} \,, \; F_{tz}=\Phi_{,z} \,.
\end{equation}
The last of the field equations need not be considered as it is satisfied automatically due to conservation laws and the other three field equations. These three have to be solved together with the Maxwell equations which in the electrostatic case have only one non-trivial component
\begin{equation}
  \Phi_{,\rho\rho}+\frac{\Phi_{,\rho}}{\rho}+\Phi_{,zz}
    =2\nu_{,\rho}\Phi_{,\rho}+2\nu_{,z}\Phi_{,z} \,.
\end{equation}
It is also easy to find
\begin{equation}  \label{FF,Weyl}
  F_{\mu\nu}F^{\mu\nu}=-2e^{-2\lambda}\left[(\Phi_{,\rho})^2+(\Phi_{,z})^2\right]
\end{equation}
as well as to check that
$4\pi(T^\rho_\rho+T^z_z)=F^{\rho z}F_{\rho z}-F^{t\phi}F_{t\phi}$
is really zero, as required for the Weyl form of the metric.

In a more restricted sense, the {\em Weyl} solutions are only those for which the gravitational potential $\nu$ and the electrostatic potential $\Phi$ are functionally dependent. As shown by \cite{Weyl-17,Majumdar-47}, if the space-time is to be asymptotically flat, the only type of such dependence allowed by the field equations is
\begin{equation}
  e^{2\nu}=1-\frac{2M}{Q}\,\Phi+\Phi^2,
\end{equation}
where $M$ and $Q$ represent total mass and charge. With such a relation, the search for $\nu$ can be reduced to a solution of Laplace equation like in the vacuum case (e.g. \cite{CooperstockC-79}). Almost all static axisymmetric electro-vacuum solutions with acceptable interpretation fall into the Weyl class; this also applies to the Majumdar--Papapetrou metrics (specified by $Q^2=M^2$) which will be treated in section \ref{MP}.

Just to remind, in case of the (generic) Weyl metric (\ref{Weyl-metric}) and electro-vacuum field equations (\ref{Laplace,nu,elstat}), (\ref{Weyl-lambda,rho,elstat}), (\ref{Weyl-lambda,z,elstat}) and (\ref{Weyl,eq4,elstat}), the Riemann tensor has non-zero components
\begin{align}
  {R^z}_{\rho z\rho}= &\;
    (\nu_{,\rho})^2+(\nu_{,z})^2-\frac{\nu_{,\rho}}{\rho} \;, \label{R2121,Weyl} \\
  {R^\phi}_{\rho\phi\rho}= &
    -\nu_{,zz}-2(\nu_{,z})^2+(\nu_{,\rho})^2 \nonumber \\ &
    -\rho\nu_{,\rho}\left[(\nu_{,\rho})^2-3(\nu_{,z})^2\right] \nonumber\\ &
    +e^{-2\nu}\rho\nu_{,\rho}\left[(\Phi_{,\rho})^2-(\Phi_{,z})^2\right] \nonumber \\ &
    -2e^{-2\nu}\rho\nu_{,z}\Phi_{,\rho}\Phi_{,z}
    +2e^{-2\nu}(\Phi_{,z})^2 \;, \\
  {R^t}_{\rho t\rho}= &
    -\nu_{,\rho\rho}-2(\nu_{,\rho})^2+(\nu_{,z})^2 \nonumber \\ &
    +\rho\nu_{,\rho}\left[(\nu_{,\rho})^2-3(\nu_{,z})^2\right] \nonumber\\ &
    -e^{-2\nu}\rho\nu_{,\rho}\left[(\Phi_{,\rho})^2-(\Phi_{,z})^2\right] \nonumber \\ &
    +2e^{-2\nu}\rho\nu_{,z}\Phi_{,\rho}\Phi_{,z} \;, \\
  {R^\phi}_{\rho\phi z}= &\;
    \nu_{,\rho z}+3\nu_{,\rho}\nu_{,z}
    +\rho\nu_{,z}\left[(\nu_{,z})^2-3(\nu_{,\rho})^2\right] \nonumber\\ &
    +e^{-2\nu}\rho\nu_{,z}\left[(\Phi_{,\rho})^2-(\Phi_{,z})^2\right] \nonumber \\ &
    -2e^{-2\nu}(1-\rho\nu_{,\rho})\,\Phi_{,\rho}\Phi_{,z} \;, \\
  {R^t}_{\phi t\phi}= &\;
    \rho^2 e^{-2\lambda}{R^z}_{\rho z\rho} \;, \\
  {R^t}_{ztz}= &\;
    {R^\phi}_{\rho\phi\rho}-2e^{-2\nu}(\Phi_{,z})^2 \;, \\
  {R^\phi}_{z\phi z}= &\;
    {R^t}_{\rho t\rho}+2e^{-2\nu}(\Phi_{,\rho})^2 \;, \\
  {R^t}_{\rho tz}= &\;
    -{R^\phi}_{\rho\phi z}-2e^{-2\nu}\Phi_{,\rho}\Phi_{,z} \;, \label{R0102,Weyl}
\end{align}
and non-zero components of the Ricci tensor simplify to
\begin{align}
  -R^t_t=R^\phi_\phi
    &= e^{-2\lambda}\left[(\Phi_{,\rho})^2+(\Phi_{,z})^2\right], \label{Ricci,Weyl,tt} \\
  -R^\rho_\rho=R^z_z
    &= e^{-2\lambda}\left[(\Phi_{,\rho})^2-(\Phi_{,z})^2\right], \\
  \qquad R_{\rho z} &= -2e^{-2\nu}\Phi_{,\rho}\Phi_{,z} \;. \label{Ricci,Weyl,rhoz}
\end{align}

In case of the {\em vacuum} Weyl metric ($\Phi\!=\!0$), the Kretschmann scalar reduces to \cite{GautreauA-67}
\begin{align}
  &R_{\mu\nu\kappa\lambda}R^{\mu\nu\kappa\lambda}=  \nonumber \\
  &=8e^{4\nu-4\lambda}
    \left[({R^\rho}_{z\rho z})^2\!+\!({R^\phi}_{\rho\phi\rho})^2\!+\!
          ({R^t}_{\rho t\rho})^2\!+\!2({R^\phi}_{\rho\phi z})^2\right],  \label{RR}
\end{align}
where the relevant components read
\begin{align}
  {R^z}_{\rho z\rho}&=
    (\nu_{,\rho})^2+(\nu_{,z})^2-\frac{\nu_{,\rho}}{\rho} \;, \\
  {R^\phi}_{\rho\phi\rho}&=
    -\nu_{,zz}\!-\!2(\nu_{,z})^2\!+\!(\nu_{,\rho})^2
    \!-\!\rho\nu_{,\rho}\left[(\nu_{,\rho})^2\!-\!3(\nu_{,z})^2\right], \\
  {R^t}_{\rho t\rho}&=
    -\nu_{,\rho\rho}\!-\!2(\nu_{,\rho})^2\!+\!(\nu_{,z})^2
    \!+\!\rho\nu_{,\rho}\left[(\nu_{,\rho})^2\!-\!3(\nu_{,z})^2\right], \\
  {R^\phi}_{\rho\phi z}&=
    \nu_{,\rho z}+3\nu_{,\rho}\nu_{,z}
    +\rho\nu_{,z}\left[(\nu_{,z})^2-3(\nu_{,\rho})^2\right].
\end{align}
It is thus clear that the scalar is nowhere negative in vacuum static axisymmetric regions.\footnote
{We will see in the following paper \cite{BasovnikS-16} that in space-times containing black holes this actually holds {\em above} horizons only, while in dynamical regions inside the holes the Kretschmann scalar can become negative. It is consistent with the given formula since below horizon the Weyl radius is imaginary effectively.}
Explicit result is
\begin{align}
  &\frac{e^{4\lambda-4\nu}}{16}\;
  R_{\mu\nu\kappa\lambda}R^{\mu\nu\kappa\lambda}=  \nonumber \\
  &=(\nu_{,\rho\rho})^2+(\nu_{,zz})^2+(\nu_{,\rho z})^2
      +\nu_{,\rho\rho}\nu_{,zz}+ \nonumber \\
  &~~ +3(1\!-\!\rho\nu_{,\rho})\left[(\nu_{,\rho})^2\!+\!(\nu_{,z})^2\right]^2
      \!+\!\rho^2\left[(\nu_{,\rho})^2\!+\!(\nu_{,z})^2\right]^3\!+ \nonumber \\
  &~~ +3\nu_{,\rho\rho}(\nu_{,\rho})^2+3\nu_{,zz}(\nu_{,z})^2
      +6\nu_{,\rho z}\nu_{,\rho}\nu_{,z}+ \nonumber \\
  &~~ +\rho\nu_{,\rho}\left[3(\nu_{,z})^2-(\nu_{,\rho})^2\right]
       (\nu_{,\rho\rho}-\nu_{,zz})+ \nonumber \\
  &~~ +2\rho\nu_{,\rho z}\nu_{,z}\left[(\nu_{,z})^2-3(\nu_{,\rho})^2\right].
  \label{Kretschmann-static-Weyl}
\end{align}

From the usual decomposition of the Riemann tensor into the Weyl tensor and contributions from the Ricci tensor and scalar curvature, one has the generally valid decomposition of the Kretschmann scalar \cite{CherubiniBCR-02}
\begin{equation}
  K=W+2R_{\mu\nu}R^{\mu\nu}-\frac{R^2}{3} \;,
\end{equation}
where $W\equiv C_{\mu\nu\kappa\lambda}C^{\mu\nu\kappa\lambda}$ is the analogous quadratic scalar given by the Weyl tensor.\footnote
{The dual scalars are equal,
\[{^*\!K}\equiv{^*\!R}_{\mu\nu\kappa\lambda}R^{\mu\nu\kappa\lambda}
  ={^*\!C}_{\mu\nu\kappa\lambda}C^{\mu\nu\kappa\lambda}\equiv{^*\!W}.\]}
For any Einstein--Maxwell space-time (pure electro-vacuum) this reduces to
\begin{align}
  K &=W+2R_{\mu\nu}R^{\mu\nu}  \nonumber \\
    &=W+2(F_{\mu\nu}F^{\mu\nu})^2+2(F_{\mu\nu}{^*\!F}^{\mu\nu})^2 \,.
\end{align}
Abdolrahimi et al. showed in \cite{AbdolrahimiFS-09} that on any static black-hole horizon the above scalars are related in a quite simple way to the Gauss curvature ${^{(2)}\!}R/2$ of the horizon's $t={\rm const}$ section (${^{(2)}\!}R$ is the Ricci scalar of the 2D horizon),\footnote
{The index `H' indicates equations only valid at the horizon.}
\begin{equation}
  W\eqH 3\,({^{(2)}\!}R-F_{\mu\nu}F^{\mu\nu})^2 \,.
\end{equation}
In the vacuum limit ($F_{\mu\nu}=0$, $C_{\mu\nu\kappa\lambda}=R_{\mu\nu\kappa\lambda}$) it reduces to $W=K\eqH 3\,({^{(2)}\!}R)^2$ which had already been presented in \cite{FrolovS-07}. This helps intuition by saying that space-time is strongly curved around places where the horizon is sharply bent.

\section{Curvature of Kerr and Kerr--Newman space-times}
\label{Kerr}

We start the discussion of specific space-times from Kerr solution, though it describes purely vacuum field of an {\em isolated} black hole (or naked singularity) and though the behaviour of its curvature scalars is quite well known. Namely, the scalars have quite complex shape in central regions, involving several sectors of negative value and non-trivial divergence at the ring singularity. It will be interesting to compare the deformation induced by rotational dragging and ring-like singularity with that caused by additional sources in the black-hole neighbourhood which will be treated in following sections.

In the Boyer--Lindquist coordinates $r$, $\theta$, the Kretschmann scalar of the Kerr solution characterized by mass $M$ and specific angular momentum $a$ is given by the surprisingly simple expression
\begin{align}
  K &\equiv R_{\mu\nu\kappa\lambda}R^{\mu\nu\kappa\lambda}=  \nonumber \\
    &=\frac{48M^2}{\Sigma^6}\,
      (r^2-a^2\cos^2\theta)(\Sigma^2-16r^2 a^2\cos^2\theta) \,,
\end{align}
where $\Sigma\equiv r^2+a^2\cos^2\theta$ is the function whose zero identifies the singularity.
First, the expression contains only even powers of all the quantities, in particular, it does not depend on the sign of $r$.
Zeros lie -- within any meridional section -- on 3+3 circles (e.g. \cite{Lake-03}),
\[r=\pm a\cos\theta, \,
  r=\pm (2-\sqrt{3})\,a\cos\theta, \,
  r=\pm (2+\sqrt{3})\,a\cos\theta\]
which are all tangent to each other at $r=0$.
In the equatorial plane the scalar is {\em independent} of $a$, namely $K(\cos\theta\!=\!0)=48M^2/r^6$, so at given $r$ it is the same as for the Schwarzschild field. On the rotation axis,
\[K(\cos^2\theta\!=\!1)=\frac{48M^2}{(r^2+a^2)^6}\,(r^2-a^2)\left[(r^2+a^2)^2-16r^2 a^2\right]\]
which is much more complicated.
This starts from a negative value $(-48M^2/a^6)$ at $r=0$, in the interval $(2-\sqrt{3})a<r<a$ it is positive, but then at $a<r<(2+\sqrt{3})a$ it falls below zero again; finally, above $r=(2+\sqrt{3})a$ it remains positive already, falling off as $1/r^6$ at infinity. If $a>M/2$, then $(2+\sqrt{3})a$ is bigger than the outer-horizon radius $r_+=M+\sqrt{M^2-a^2}$, so for moderately and rapidly spinning holes the invariant is negative along the axis even {\em above the horizon}, up to $r=(2+\sqrt{3})a$.

The Chern--Pontryagin scalar comes out quite simple as well,
\begin{align}
  {^*\!K} &\equiv {^*\!R}_{\mu\nu\kappa\lambda}R^{\mu\nu\kappa\lambda}=  \nonumber \\
          &=\frac{96M^2}{\Sigma^6}\,ra\cos\theta\,
            (3r^2-a^2\cos^2\theta)(r^2-3a^2\cos^2\theta) \,.
\end{align}
This is exactly opposite at $r>0$ and $r<0$ sheets as well as on opposite sides from the equatorial plane.
Zeros lie on 2+2 circles
\[r=\pm\frac{a}{\sqrt{3}}\,\cos\theta, \quad
  r=\pm\sqrt{3}\,a\cos\theta \,,\]
again tangent to each other at $r=0$, and also on $r=0$ and in the whole equatorial plane. On the axis one has
\[{^*\!K}(\cos\theta\!=\!\pm 1)=\pm\frac{96M^2}{(r^2+a^2)^6}\,ra\,(3r^2-a^2)(r^2-3a^2)\,;\]
this grows toward positive values when going from zero at $r=0$ toward positive radii, then at $a/\sqrt{3}<r<\sqrt{3}\,a$ it is negative and finally positive again above $r=\sqrt{3}\,a$, falling off as $1/r^7$ at infinity.

\begin{figure*}
\centering
\includegraphics[width=0.85\textwidth]{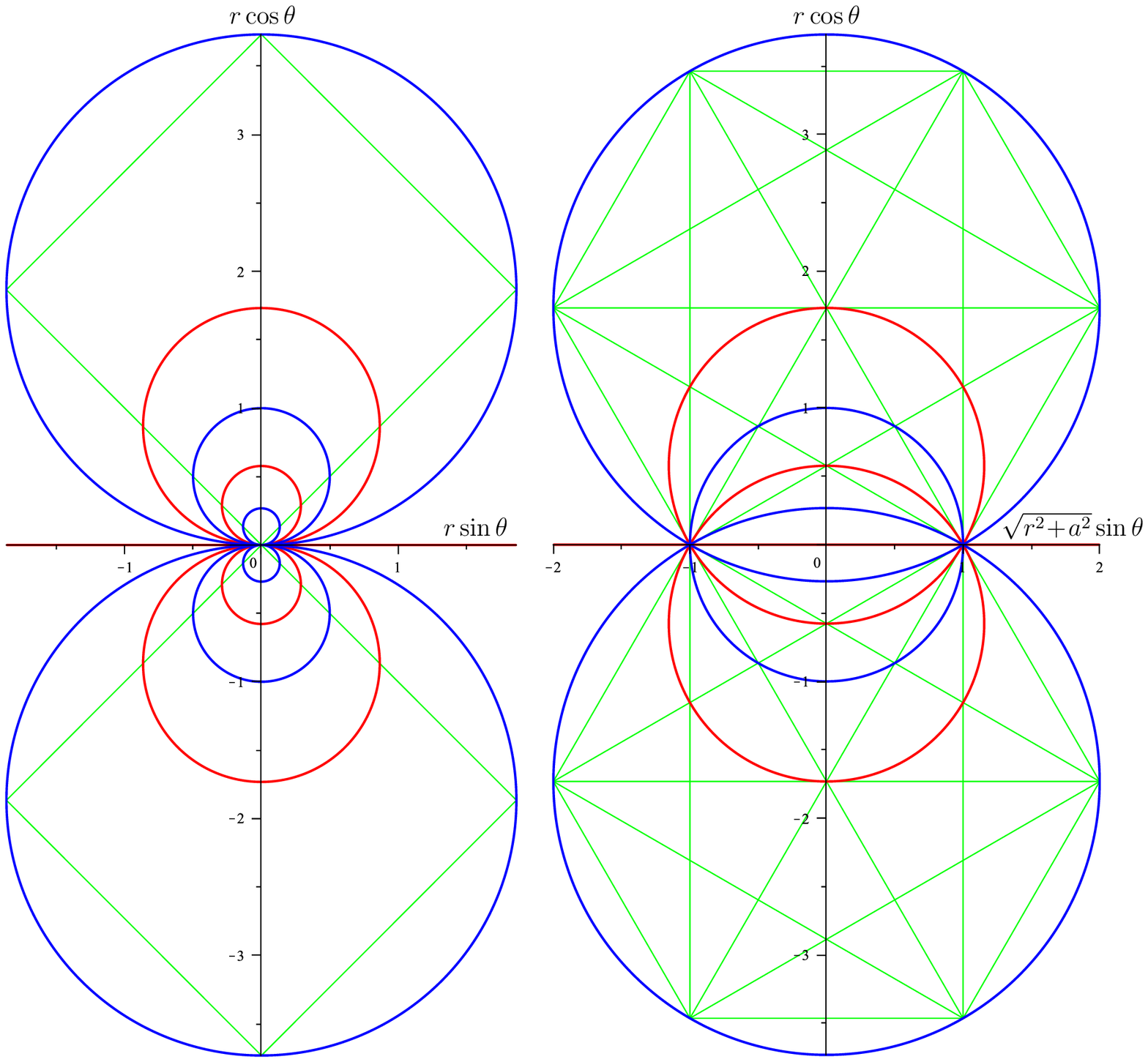}
\caption
{Curvature of the Kerr space-time represented in the Boyer--Lindquist coordinates $r\sin\theta$, $r\cos\theta$ ({\bf left}) and in the Kerr--Schild coordinates $R=\sqrt{r^2+a^2}\,\sin\theta$, $Z=r\cos\theta$ ({\bf right}). The six/three blue circles indicate zeros of the Kretschmann scalar and the four/two red circles (plus red-coloured horizontal axis) indicate zeros of the dual, Chern--Pontryagin scalar. The arrangement is quite ``miraculous" in the Kerr--Schild plot: all the circles intersect at the singularity $(R=a,z=0)$ and define a ($\pi/6$)-segmentation of meridional planes there; a remarkable symmetry of the pattern is revealed on tangents to the circles drawn (in green colour) at the singularity (note, for example, that the tangents only intersect at the circles). In the Boyer--Lindquist picture (left), the pattern based on circles' tangents is of course degenerate and the only other straight lines one can draw are diagonals crossing the circles' at their leftmost/rightmost points.}
\label{Kerr-scalars}
\end{figure*}

The scalars seem to prove very complex (though highly symmetrical) shape of space-time fabric in the central Kerr region (see figure \ref{Kerr-scalars}), but the modulus of the complex number $K-{\rm i}\,{^*\!K}$ comes out extremely simple,\footnote
{From treatment of the Petrov-type-$D$ metrics in the Newman--Penrose formalism it is known that $W-{\rm i}\,{^*\!W}=48(\Psi_2)^2$, where $\Psi_2$ is the second NP-tetrad projection of the Weyl tensor. In the vacuum case it is $W\!=\!K$ and for the Kerr metric one has $\Psi_2=-M/(r-{\rm i}\,a\cos\theta)^3$, from where the modulus $|K-{\rm i}\,{^*\!K}|$ follows immediately.}
\begin{equation}
  |K-{\rm i}\,{^*\!K}|\equiv\sqrt{K^2+{^*\!K}^2}=\frac{48M^2}{\Sigma^3} \,.
\end{equation}
Hence, if {\em both} the independent quadratic curvature scalars are combined in an obvious manner, they give exactly the same message as for the Schwarzschild field, only the singularity is now given by $\Sigma=0$ instead of $r=0$. In particular, the quadratic curvature does {\em not} indicate any directional behaviour of the Kerr singularity, as already pointed out by \cite{Lake-03}. Both $r>0$ and $r<0$ sheets of the metric have the same curvature structure (just with ${^*\!K}$ having opposite sign), which is in contrast with causal structure, very different in the two sheets.

The two Kerr-field curvature scalars are represented even more neatly in the Kerr--Schild coordinates $R=\sqrt{r^2+a^2}\,\sin\theta$, $Z=r\cos\theta$. The Kretschmann scalar reads
\begin{equation}
  K=\frac{48M^2}{\Sigma^6}\,(R^2+Z^2-a^2)\left[(R^2+Z^2-a^2)^2-12a^2 Z^2\right],
\end{equation}
where $\Sigma^2=(R^2+Z^2-a^2)^2+4a^2 Z^2$.
Hence \cite{CherubiniBCR-02,Lake-03}, in the $(R,Z)$ plane, it is zero and changes sign on 3 circles given by
\begin{equation}
  R^2+Z^2=a^2
  \quad {\rm and} \quad
  R^2+(Z\pm\sqrt{3}\,a)^2=4a^2 \,.
\end{equation}
These circles intersect at the singularity $(R=a,Z=0)$ exactly under the angles $\pi/3$.
The Chern--Pontryagin scalar assumes the form
\begin{equation}
  {^*\!K}=\frac{96M^2 aZ}{\Sigma^6}\left[3(R^2+Z^2-a^2)^2-4a^2 Z^2\right],
\end{equation}
so it vanishes and changes sign on 2 circles
\begin{equation}
  R^2+\left(Z\pm\frac{a}{\sqrt{3}}\right)^2=\frac{4a^2}{3} \,.
\end{equation}
The circles intersect at the singularity as well, again forming (together with the $Z=0$ axis) a $(\pi/3)$-segmentation of the meridional plane around the singularity which is exactly complementary to the one defined by zero circles of the Kretschmann scalar. This makes the whole pattern quite ``magic". In particular, the crossing under $60^\circ$ means that the circles of Chern--Pontryagin-scalar zeros go through each other's centre, while in their outer parts they pass exactly through centres of the Kretschmann-scalar big two circles. Tangents to the circles drawn at the singularity form a highly symmetrical triangular pattern inscribed to the circles, see figure \ref{Kerr-scalars} (right plot).

\begin{figure*}
\centering
\includegraphics[width=\textwidth]{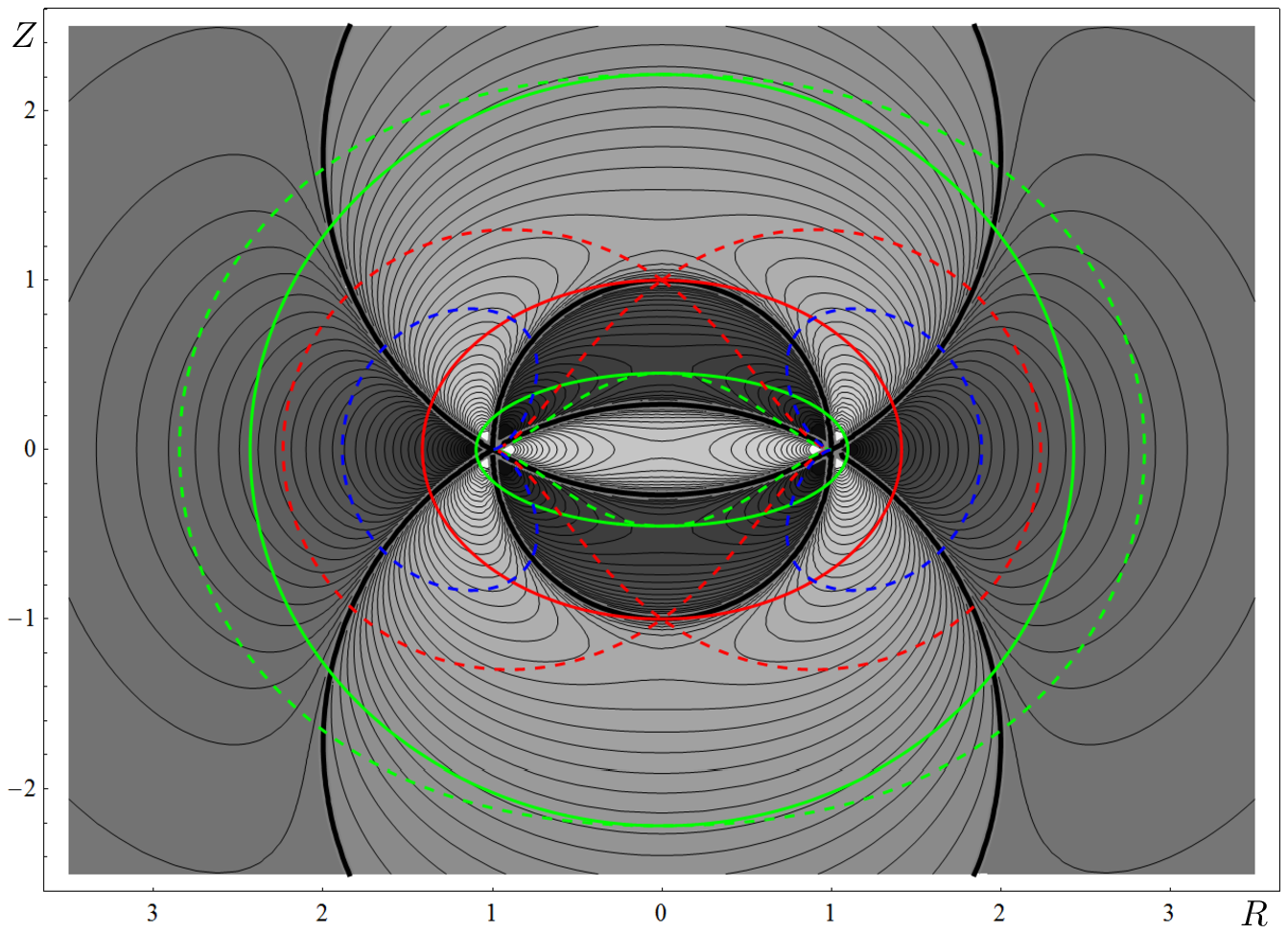}
\caption
{Kretschmann scalar $K$ in the central part of the Kerr space-time, as depicted in the meridional plane represented in the Kerr--Schild coordinates $(R,Z)$, with ergospheres indicated for three different values of $a/M$ -- $0.8$ (green), $1.0$ (red: extreme) and $1.\overline{33}$ (blue: naked). The axes are given in the units of $a$, which makes the curvature pattern unchanging, while the horizons (drawn in solid lines) and static-limit surfaces (drawn in short-dashed lines) shift accordingly with $a/M$. The contours of $K$ are shown together with grey shading which indicates its value: darker/lighter grey means bigger positive/negative value; $K=0$ circles are drawn in solid black.}
\label{Kerr-Kretschmann}
\end{figure*}

In both scalars the mass $M$ only scales the multiplicative factor, and the dependence on the other parameter $a$ also reduces, in the Kerr--Schild coordinates, to a simple scaling; the ``curvature pattern" is independent. Actually, if the scalars are expressed in terms of dimensionless $\tilde{a}\equiv a/M$, $\tilde{R}\equiv R/M$ and $\tilde{Z}\equiv Z/M$, then they are proportional to $M^{-4}$ and (if $\chi$ is some constant)
\begin{align}
  K(M;\chi\tilde{a},\chi\tilde{R},\chi\tilde{Z})
     &= \chi^{-6}K(M;\tilde{a},\tilde{R},\tilde{Z}), \\
  {^*\!K}(M;\chi\tilde{a},\chi\tilde{R},\chi\tilde{Z})
     &= \chi^{-6}K(M;\tilde{a},\tilde{R},\tilde{Z}).
\end{align}
This implies that the curvature pattern is not correlated with the appearance and position of structures given by metric itself, like static-limit surfaces and horizons, in particular, it does not distinguish between black holes and naked singularities. However, since the dependence of the radii of static limits and horizons on $a/M$ is different and does {\em not} reduce to any simple scaling, the curvature pattern and metric features ``fit together" differently for different $a/M$ -- see figure \ref{Kerr-Kretschmann}.

Let us note that recently \cite{AbdelqaderL-13} presented a thorough picture of gradient fields of the four Kerr-metric Weyl invariants (of which two coincide with the scalars treated here), showing an interesting dependence on the centre's spin. (See \cite{Lake-12} for an introductory review on invariants polynomial in curvature and on the gradient-flow method, and \cite{AbdelqaderL-12} for its application to Chazy--Curzon solution.)
Let us also add here that an alternative way of curvature visualization has been developed by \cite{Nichols-etal-11}; it was applied to stationary black holes by \cite{Zhang-etal-12}.

\subsection{Kerr--Newman generalization}

If the centre is endowed with an electric charge $Q$, the above ``miraculously simple" picture is somewhat disturbed.
The Kretschmann-scalar expression for the Kerr--Newman space-time with parameters $M$, $a$, $Q$ remains rather simple,
\begin{align}
  K&= \frac{8}{\Sigma^6}
      \big[6M^2(r^2-a^2\cos^2\theta)(\Sigma^2-16r^2a^2\cos^2\theta) \nonumber \\
   &~ \qquad -12MQ^2r(r^4-10r^2a^2\cos^2\theta+5a^4\cos^4\theta) \nonumber \\
   &~ \qquad +Q^4(7r^4-34r^2a^2\cos^2\theta+7a^4\cos^4\theta)\big].
\end{align}
As opposed to the uncharged, Kerr case, the radius $r$ now appears in both even and odd powers, so the curvature ``landscape" is different for the $r\!>\!0$ and $r\!<\!0$ space sheets. Besides several special directions $\theta$ along which the invariant does not diverge at $\Sigma\rightarrow 0$, the only case when it remains finite at the singularity is when {\em both} $M$ and $Q$ vanish. Actually, charge $Q$ even makes the singularity {\em stronger} than mass $M$, in particular, the limit $M=0$, $a=0$, $Q\neq 0$ yields $K=56Q^4/r^8$ which is more divergent at $r\rightarrow 0$ than the Schwarzschild expression $K=48M^2/r^6$.
In the equatorial plane the dependence on $a$ is again suppressed and the scalar assumes the Reissner--Nordstr\"om form
$K(\cos\theta\!=\!0)=(8/r^8)\,(6M^2r^2-12MQ^2r+7Q^4)$. (See \cite{Henry-00} for visualisation.)

The other independent scalar reads
\begin{align}
  &\!\!\!\!{^*\!K}= \frac{96a\cos\theta}{\Sigma^6}\;
                    (3Mr^2-Ma^2\cos^2\theta-2Q^2 r)\,\times \nonumber \\
         &\quad \times
            \left[Mr(r^2-3a^2\cos^2\theta)-Q^2(r^2-a^2\cos^2\theta)\right],
\end{align}
so it also depends on the sign of $r$ non-trivially. This even applies to the modulus of $K-{\rm i}\,{^*\!K}$ which is no longer that short as in the Kerr limit,
\begin{align}
  K^2&+{^*\!K}^2=\frac{64}{\Sigma^{10}}
                 \left[\Sigma^2(6M^2\Sigma-12MrQ^2+7Q^4)^2- \right. \nonumber \\
               & \left. {} -24Q^4 a^2\cos^2\theta\,(3Mr^2-Ma^2\cos^2\theta-2Q^2 r)^2\right].
\end{align}

\begin{figure*}
\centering
\includegraphics[width=0.7\textwidth]{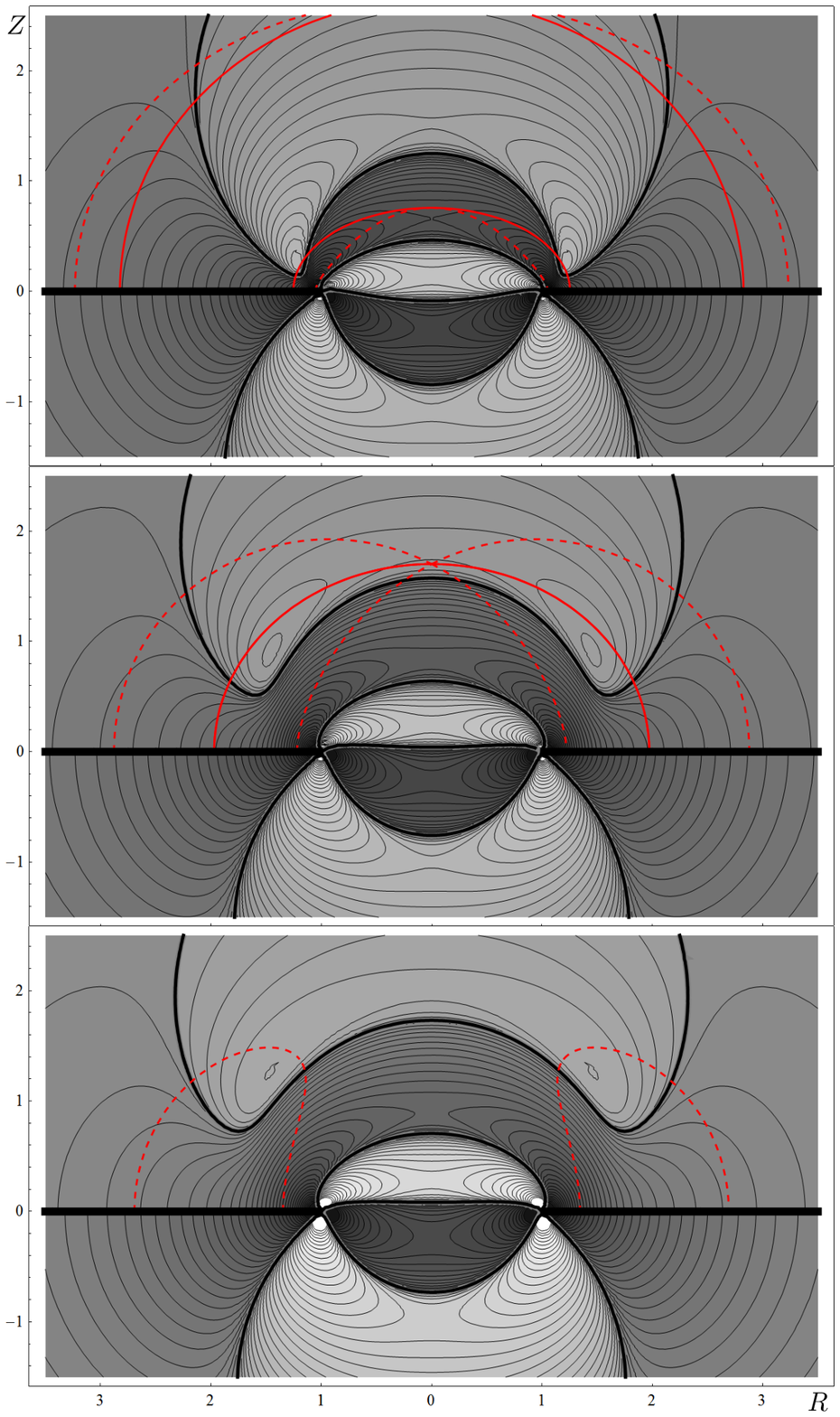}
\caption
{Kretschmann-scalar ``landscape" in the central part of the Kerr--Newman space-time, depicted in the Kerr--Schild coordinates $(R,Z)$ like in figure \ref{Kerr-Kretschmann} (the Kerr case), for $a\!=\!(10/17)M\!\doteq\!0.59M$ and three different charges $Q$ (from top to bottom): $M$, $(\sqrt{189}/17)M\!\doteq\!0.81M$ (extreme) and $(15/17)M\!\doteq\!0.88M$ (naked).
The axes are given in the units of $a$, solid red lines represent horizons and short-dashed red lines represent static limits. Upper/lower parts of the plots show $r\!>\!0$/$r\!<\!0$ space sheets, with discontinuity in the equatorial plane indicated.}
\label{KN-Kretschmann}
\end{figure*}

With growing charge $Q$, the pattern of scalars gradually moves away from the Kerr picture, namely in the Kerr--Schild coordinate representation the zero-value circles ``reconnect" and the disconnected negative-value regions get kidney-shaped -- see figure \ref{KN-Kretschmann}.

\section{Majumdar--Papapetrou binary black hole}
\label{MP}

In order to subject a black hole to a strong and highly inhomogeneous external field, the best possibility seems to be an another black hole. The resulting binary would almost never be in stationary equilibrium, which the Einstein equations ingeniously ``repair" by adding a singular struts into a system. The only known stationary (actually even static) regular possibility is the Majumdar--Papapetrou configuration when the gravitational attraction between sources is counter-balanced by electric repulsion. Exact equilibrium requires, like in the case of Newtonian gravity, that all the sources have charges of the same sign and of extreme values equal to their masses (see e.g. \cite{Majumdar-47,HartleH-72}). In this section we will consider a binary version of these solutions: two extreme black holes of masses $M_1$, $M_2$ and charges $Q_1=\pm M_1$, $Q_2=\pm M_2$ at some coordinate distance $2b$ in a static equilibrium. Such a system is axially symmetric about axis going through the black-hole centres; it is reflectionally symmetric only if the masses are equal. The solution is electro-vacuum, so its metric can be written in the Weyl form (\ref{Weyl-metric}); more specifically, it belongs to the Weyl class with the relation $e^{2\nu}=(1\mp\Phi)^2$ between the gravitational and electrostatic potentials.

\subsection{Metric and coordinates}

The Majumdar--Papapetrou family of solutions provides the only known case of singularity-free stationary electrovacuum space-times with more than one black hole \cite{ChruscielCH-12}. Its metric is usually presented in Cartesian-type coordinates $(x,y,z)$,
\begin{equation}  \label{MP-metric}
  {\rm d}s^2=-N^2{\rm d}t^2+N^{-2}({\rm d}x^2+{\rm d}y^2+{\rm d}z^2) \,,
\end{equation}
where the lapse function $N\equiv e^\nu$ is given by
\[\frac{1}{N}=1+\sum_{j=1}^{n}\frac{M_j}{|\vec{r}-\vec{r}_j|} \;,\]
$n$ being the number of black holes and $M_j$ and $\vec{r}_j\equiv(x_j,y_j,z_j)$ denoting their masses and positions (namely the positions of their horizons which are represented as points in the above coordinates). The electromagnetic field is given by potential $A_\mu=(\pm N,0,0,0)$.\footnote
{One should actually take $A_\mu=(\pm(N-1),0,0,0)$ for a full consistence with the general electrostatic expression $A_\mu=(-\Phi,0,0,0)$ and with the Majumdar--Papapetrou prescription $N^2\equiv e^{2\nu}=(1\mp\Phi)^2$, but conventionally the lapse itself is chosen in the role of $\Phi$. This only corresponds to normalising the potential to $1$ instead of $0$ at spatial infinity; in particular, no difference arises in the field ($F_{\mu\nu}$).}
For just two black holes, the system is axially symmetric about their connecting line. Identifying the latter as the $z$-axis, the Weyl form of the metric follows immediately by putting $x=\rho\cos\phi$, $y=\rho\sin\phi$; since it involves $g_{\rho\rho}=g_{zz}=N^{-2}\equiv e^{-2\nu}$, it corresponds to $\lambda=0$. Let us choose the coordinate origin so that the horizons lie at $(0,0,+b)$ and $(0,0,-b)$. The lapse then reads
\begin{equation}
  \frac{1}{N}=1+\frac{M_1}{\sqrt{\rho^2+(z-b)^2}}+\frac{M_2}{\sqrt{\rho^2+(z+b)^2}} \;,
\end{equation}
where the denominators represent coordinate distances of a given location from horizons in the $(\rho,z)$ plane.
Note that the separation of black holes $2b$ must not be too small in order for the binary not to be enclosed in a common apparent horizon -- see \cite{JaramilloL-11} (table I there).

\begin{figure}
\centering
\includegraphics[width=\columnwidth]{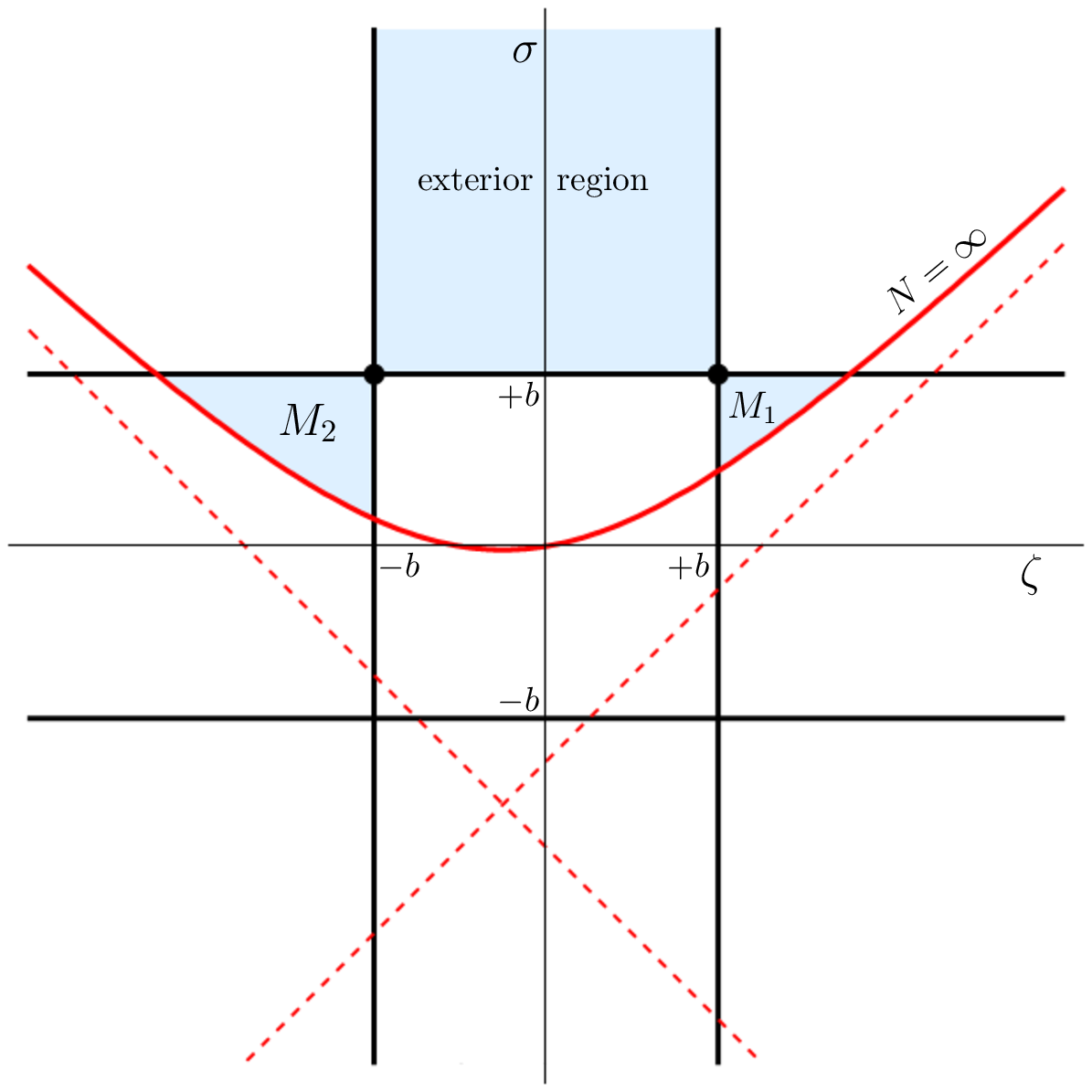}
\caption
{Meridional plane of the binary Majumdar--Papapetrou solution represented in the coordinates $(\sigma,\zeta)$. The black holes have masses $M_1$, $M_2=1.5 M_1$ and their horizons are placed at $z=\pm b=\pm M_1$ on the Weyl axis. Indicated in light blue are physical regions -- the two black holes and the exterior domain. The horizons are just points at $\sigma=b$, $\zeta=\pm b$ and the singularities inside black holes (where $N=\infty$) lie on the hyperbola given in red colour. The symmetry axis is made of 7 segments: the part between radial infinity and the 1st singularity, lying in the direction opposite to the 2nd black hole, is given by $\zeta=b$, with $\sigma>b$ above the (1st) horizon and $\sigma<b$ below it; the part between the singularities is given by $\sigma=b$, with $\zeta>b$ below the 1st horizon, $-b<\zeta<+b$ between the horizons and $\zeta<-b$ below the 2nd horizon; and the part between the 2nd singularity and radial infinity is given by $\zeta=-b$, with $\sigma<b$ below the (2nd) horizon and $\sigma>b$ above it.}
\label{sigma-zeta}
\end{figure}

The Weyl-type coordinates $(\rho,z)$ cover only the region outside the black holes (their horizons appear as points). In order to also include the inner regions, we will introduce two other coordinate couples in the meridional planes. The first of them, $(\sigma,\zeta)$, is given by
\begin{equation}  \label{rho,z->sigma,zeta}
  (\sigma-\zeta)^2=\rho^2+(z-b)^2, \quad
  (\sigma+\zeta)^2=\rho^2+(z+b)^2,
\end{equation}
or, in the inverse sense,
\begin{equation}  \label{sigma,zeta->rho,z}
  \rho^2=x^2+y^2=\frac{(\sigma^2-b^2)(b^2-\zeta^2)}{b^2} \;, \quad
  z=\frac{\sigma\zeta}{b} \;.
\end{equation}
The lapse function then appears as
\begin{equation}  \label{N,sigma-zeta}
  \frac{1}{N}=1+\frac{M_1}{\sigma-\zeta}+\frac{M_2}{\sigma+\zeta}
\end{equation}
and the Majumdar--Papapetrou metric as
\begin{align}
  {\rm d}s^2=
  &-N^2{\rm d}t^2
   +\frac{1}{N^2}\left[\frac{(\sigma^2-b^2)(b^2-\zeta^2)}{b^2}\,{\rm d}\phi^2+ {} \right. \nonumber\\
                       &\left. {}
                        +(\sigma^2-\zeta^2)
                         \left(\frac{{\rm d}\sigma^2}{\sigma^2-b^2}+
                               \frac{{\rm d}\zeta^2}{b^2-\zeta^2}\right)\right].
  \label{PM-metric-sigma,zeta}
\end{align}
Representation of the meridional plane in $(\sigma,\zeta)$ is depicted in figure \ref{sigma-zeta}.
The domain of outer communications, just covered by $(\rho,z)$, corresponds to the ranges
$\sigma\in\langle b,\infty)$, $\zeta\in\langle -b,+b\rangle$. The black-hole horizons are given by $\sigma=b$, $\zeta=\pm b$, and the dynamical regions below horizons by $\sigma<b$, $|\zeta|>b$, being bounded ``from bottom" by space-time singularities localized where $1/N=0$ and represented as parts of the hyperbola
\begin{equation}
  \left(\sigma+\frac{M_1+M_2}{2}\right)^2-\left(\zeta-\frac{M_1-M_2}{2}\right)^2=M_1 M_2 \;.
\end{equation}
The hyperbola has asymptotes $\zeta=\pm\sigma+\frac{1}{2}(M_1-M_2)$ and always passes through ($\sigma\!=\!0,\zeta\!=\!0$).

The transformation (\ref{rho,z->sigma,zeta}), (\ref{sigma,zeta->rho,z}) clearly does not recognize the signs of $x$ and $y$, but this does not matter if only meridional projection is in question (one may identify the points with all possible combinations of $x$ and $y$ signs due to the axial symmetry about the $z$-axis).

The last coordinate system we will use is the usual spheroidal system $(r,\theta)$, adapted to the ``first" black hole,
\begin{equation}  \label{spheroidal-coords}
  (r-M_1)^2=\rho^2+(z-b)^2, \quad
  \tan\theta=\frac{\rho}{z-b} \,,
\end{equation}
with the inverse relation
\begin{equation}
  \rho=(r-M_1)\,\sin\theta, \quad
  z-b=(r-M_1)\,\cos\theta.
\end{equation}
The lapse function now writes
\begin{align}
  \frac{1}{N}&= 1+\frac{M_1}{r-M_1}+\frac{M_2}{\sqrt{(r-M_1+2b\cos\theta)^2+4b^2\sin\theta^2}}
              \nonumber \\
             &= \frac{1}{1-\frac{M_1}{r}}+\frac{M_2}{\sqrt{(r-M_1+2b\cos\theta)^2+4b^2\sin\theta^2}}
             \label{N,spheroidal}
\end{align}
and the metric assumes the form
\begin{equation}  \label{metric-spheroidal}
    {\rm d}s^2=
    -N^2{\rm d}t^2+\frac{{\rm d}r^2}{N^2}
    +\frac{\left(1-\frac{M_1}{r}\right)^2}{N^2}\;
     r^2({\rm d}\theta^2+\sin^2\theta\,{\rm d}\phi^2) \;.
\end{equation}
In the limit of a single black hole ($M_2=0$), this reduces to the extreme Reissner--Nordstr\"om metric in spherical coordinates,
\begin{align}
  &M_2=0 \quad \Longrightarrow \quad N=1-\frac{M_1}{r} \nonumber \\
  &\Rightarrow \;\;
   {\rm d}s^2=
     -N^2{\rm d}t^2+\frac{{\rm d}r^2}{N^2}
     +r^2({\rm d}\theta^2+\sin^2\theta\,{\rm d}\phi^2) \;.
\end{align}
In the $(r,\theta)$ coordinates, the first black-hole horizon lies on $r=M_1$ and the second one remains point-like at $\theta=\pi$, $r=M_1+2b$. Singularity ($1/N=0$) is only reached inside the first black hole, at radius given by solution of a quartic equation.

\subsection{Gravitational acceleration and curvature}

Since the electrostatic potential is $\Phi=\pm(1-N)$ (actually the re-normalized alternative $\Phi=\mp N$ is being considered conventionally) and $\lambda=0$ for the Majumdar--Papapetrou field, the electromagnetic scalar (\ref{FF,Weyl}) is proportional to the gravitational-acceleration square (\ref{kappa2}) which for any Weyl metric reads $\kappa^2=e^{2\nu-2\lambda}\left[(N_{,\rho})^2+(N_{,z})^2\right]$:
\begin{align}
  F_{\mu\nu}F^{\mu\nu}=-2\left[(N_{,\rho})^2+(N_{,z})^2\right]
                      =-2\kappa^2/N^2 \\
  \Longrightarrow \quad
  R_{\mu\nu}R^{\mu\nu}=(F_{\mu\nu}F^{\mu\nu})^2
                      =4\kappa^4/N^4 \,.
\end{align}
Both $N$ and $\kappa$ vanish on both horizons, while the Ricci-square scalar combines them into the finite value $4/(M_1)^4$ on the first horizon and $4/(M_2)^4$ on the second horizon; $R_{\mu\nu}R^{\mu\nu}$ diverges where $N$ has a divergence (which is {\em not} at $r=0$).
Ricci tensor itself is also found immediately from its general Weyl form (\ref{Ricci,Weyl,tt})--(\ref{Ricci,Weyl,rhoz}),
\begin{align}
  -R^t_t &=R^\phi_\phi=(N_{,\rho})^2+(N_{,z})^2, \nonumber \\
  -R^\rho_\rho &=R^z_z=(N_{,\rho})^2-(N_{,z})^2, \nonumber \\
   R^\rho_z &=-2N_{,\rho}N_{,z} \;.  \label{Ricci,MP}
\end{align}

Riemann tensor of the Majumdar--Papapetrou metric can be obtained according to the general Weyl electro-vacuum form (\ref{R2121,Weyl})--(\ref{R0102,Weyl}), just using $\nu_{,i}=-N\,(1/N)_{,i}\,$, but it will be better to evaluate it in some coordinates which also cover the black-hole interior. We choose the $(\sigma,\zeta)$ coordinates in which the tensor appears relatively simple, mainly if expressed suitably in terms of sums and differences of its mixed components,
\begin{align*}
  &{R^{t\phi}}_{t\phi}={R^{\sigma\zeta}}_{\sigma\zeta}= \\
     &= -N^4\left[\frac{M_1}{(\sigma-\zeta)^3}+\frac{M_2}{(\sigma+\zeta)^3}
                      +\frac{4M_1 M_2}{(\sigma^2-\zeta^2)^3}\,b^2\right], \\
  &{R^{\phi\sigma}}_{\phi\zeta}-{R^{t\sigma}}_{t\zeta}= \\
     &= 6N^4\,\frac{\sigma^2-b^2}{\sigma^2-\zeta^2}
        \left[\frac{M_1}{(\sigma-\zeta)^3}-\frac{M_2}{(\sigma+\zeta)^3}
              +\frac{4M_1 M_2}{(\sigma^2-\zeta^2)^3}\,\sigma\zeta\right], \\
  &{R^{\phi\sigma}}_{\phi\zeta}+{R^{t\sigma}}_{t\zeta}
     = 2N^4\,\frac{\sigma^2-b^2}{\sigma^2-\zeta^2}
        \left[\frac{(M_1)^2}{(\sigma-\zeta)^4}-\frac{(M_2)^2}{(\sigma+\zeta)^4}\right], \\
  &{R^{\phi\sigma}}_{\phi\sigma}-{R^{t\zeta}}_{t\zeta}
     = 2N^4\,\frac{b^2-\zeta^2}{\sigma^2-\zeta^2}
        \left[\frac{M_1}{(\sigma-\zeta)^2}-\frac{M_2}{(\sigma+\zeta)^2}\right]^2, \\
  &{R^{\phi\zeta}}_{\phi\zeta}-{R^{t\sigma}}_{t\sigma}
     = 2N^4\,\frac{\sigma^2-b^2}{\sigma^2-\zeta^2}
        \left[\frac{M_1}{(\sigma-\zeta)^2}+\frac{M_2}{(\sigma+\zeta)^2}\right]^2, \\
  &{R^{\phi\sigma}}_{\phi\sigma}+{R^{t\zeta}}_{t\zeta}= \\
     &= \frac{-2N^4}{\sigma^2-\zeta^2}
        \left\{\left[\frac{M_1}{(\sigma-\zeta)^3}+\frac{M_2}{(\sigma+\zeta)^3}\right]
               (\sigma^2+2\zeta^2-3b^2)+
               {} \right. \nonumber \\ & \left. {} \qquad\qquad\quad
              +\frac{4M_1 M_2}{(\sigma^2-\zeta^2)^3}
               (3\sigma^2\zeta^2-b^2\zeta^2-2b^2\sigma^2)\right\}, \\
  &{R^{\phi\zeta}}_{\phi\zeta}+{R^{t\sigma}}_{t\sigma}= \\
     &= \frac{2N^4}{\sigma^2-\zeta^2}
        \left\{\left[\frac{M_1}{(\sigma-\zeta)^3}+\frac{M_2}{(\sigma+\zeta)^3}\right]
               (2\sigma^2+\zeta^2-3b^2)+
               {} \right. \nonumber \\ & \left. {} \qquad\qquad\quad
              +\frac{4M_1 M_2}{(\sigma^2-\zeta^2)^3}
               (3\sigma^2\zeta^2-2b^2\zeta^2-b^2\sigma^2)\right\}.
\end{align*}
The Kretschmann scalar can be written fully explicitely,
\begin{align}
  &\frac{(\sigma^2-\zeta^2)^8}{8N^8}\;K= \\
  & (M_1)^4(\sigma+\zeta)^8+(M_2)^4(\sigma-\zeta)^8+ {} \nonumber \\
  & +4M_1 M_2
     \left[(M_1)^2(\sigma+\zeta)^4+(M_2)^2(\sigma-\zeta)^4\right] \times \nonumber \\
     &\qquad\qquad\qquad \times {}
     (\sigma^2-\zeta^2)(\sigma^2+\zeta^2-2b^2)+ {} \nonumber \\
  & +2(M_1 M_2)^2
     \left[3(\sigma^2\!-\zeta^2)^2-8(\sigma^2\!-b^2)(b^2\!-\zeta^2)+48b^4\right] \times \nonumber \\
     &\qquad\qquad\qquad \times {}
      (\sigma^2-\zeta^2)^2+ {} \nonumber \\
  & +24(M_1)^2 M_2
     \left[3(\sigma\zeta+b^2)^2-b^2(\sigma+\zeta)^2\right] \times \nonumber \\
     &\qquad\qquad\qquad \times {}
     (\sigma^2-\zeta^2)^2(\sigma+\zeta)+ {} \nonumber \\
  & +24M_1(M_2)^2
     \left[3(\sigma\zeta-b^2)^2-b^2(\sigma-\zeta)^2\right] \times \nonumber \\
     &\qquad\qquad\qquad \times {}
     (\sigma^2-\zeta^2)^2(\sigma-\zeta)+ {} \nonumber \\
  & +6\left[(M_1)^2(\sigma+\zeta)^6+(M_2)^2(\sigma-\zeta)^6\right]
     (\sigma^2-\zeta^2)^2+ {} \nonumber \\
  & +12M_1 M_2
     \left[(\sigma^2+\zeta^2-2b^2)^2-2(\sigma^2-b^2)(b^2-\zeta^2)\right] \times \nonumber \\
     &\qquad\qquad\qquad \times {}
     (\sigma^2-\zeta^2)^3 \,. \nonumber
\end{align}
On the first horizon ($\sigma=b$, $\zeta=+b$) it yields $8/(M_1)^4$ and on the second horizon ($\sigma=b$, $\zeta=-b$) it yields $8/(M_2)^4$. It is easy to check that the formula \cite{AbdolrahimiFS-09}
\begin{equation}
  K\eqH 3\,({^{(2)}\!}R-F_{\mu\nu}F^{\mu\nu})^2+2\,(F_{\mu\nu}F^{\mu\nu})^2 \,,
\end{equation}
valid on any static electro-vacuum horizon, really holds for the Majumdar--Papapetrou values: the Ricci scalars of our horizon 2D metrics\footnote
{This is in fact only the metric of the first horizon: in the coordinates adapted to the first black hole, the second horizon is singular.}
\begin{equation}
  {\rm d}s^2 \eqH
     (M_1)^2({\rm d}\theta^2+\sin^2\theta\,{\rm d}\phi^2)
\end{equation}
are ${^{(2)}\!}R\eqH 2/(M_1)^2$ and ${^{(2)}\!}R\eqH 2/(M_2)^2$ respectively for the first and the second horizon, and the electromagnetic invariant assumes exactly the same values there, namely $F_{\mu\nu}F^{\mu\nu}\eqH {^{(2)}\!}R$ holds on both horizons, so the above formula reduces to
\begin{equation}
  K\eqH 2\,(F_{\mu\nu}F^{\mu\nu})^2\eqH 2{^{(2)}\!}R^2
   \eqH 8/(M_{1,2})^4 \,.
\end{equation}

\subsection{Meissner-like effect and geometry of the horizon}

It is well known that rotating and charged black holes tend to ``expel" stationary axisymmetric external fields; as the black hole approaches the extreme state, the external field-lines are pushed out and their flux across any part of the horizon vanishes. This was demonstrated on external (electro)magnetic fields, either test (or weak) ones in the Kerr \cite{Wald-74,KingLK-75,BicakJ-85,Karas-89} and the Reissner--Nordstr\"om \cite{BicakD-80,BiniGR-08} space-times, or within the exact Ernst solution for magnetized Kerr(--Newman) black holes \cite{Karas-88,AlievG-89,KarasV-91,KarasB-00}; see \cite{BicakL-00} for a review. Astrophysical implications of the effect are still under discussion, in particular, it has been shown \cite{KomissarovM-07,Takamori-etal-11} that the fields {\em can} penetrate the horizon if currents are present. In \cite{Semerak-02-cjp} we studied a stationary and axisymmetric exact solution \cite{ZellerinS-00,Semerak-02-cqg} describing a rotating black hole in an external gravitational field generated by a disc and found that in the extreme limit the ``external" field vanishes on the black-hole horizon.

However, it is not clear whether the ``gravitational Meissner effect" has actually a good sense. Namely, there are two simple coordinate-independent ways how to characterize the gravitational-field intensity: by a magnitude of four-acceleration of some fiducial (``stationary") observer, or by some invariant determined by first derivatives of metric. The first proposal runs into problems exactly on the horizon, because acceleration of a stationary observer is infinite there in any case. The second proposal leads to the gravitational acceleration $\kappa$; this is in fact a renormalized (by lapse $N$) version of the stationary-observer acceleration. It stays regular on the horizon (being called {\it surface gravity} there), but on extreme horizons it is {\em zero by definition}. In this sense, extreme holes expel (all) gravitational fields {\em by definition}, so the ``gravitational Meissner" effect occurs {\em inavitably}.

The gravitational acceleration $\kappa$ really vanishes on the horizons of our Majumdar--Papapetrou binary black hole, and the scalar obtained from it by further differentiation,
\[g^{\alpha\beta}\kappa_{,\alpha}\kappa_{,\beta}=
  N^2\left[(\kappa_{,1})^2+(\kappa_{,2})^2\right],\]
is zero there as well.
One can still analyse several other geometric quantities which remain finite at the horizon. Firstly, it is the ratio $\kappa/N$ that goes to
\begin{equation}
  \lim_{N\rightarrow 0}\frac{\kappa}{N}=\frac{1}{M_1} \;.
\end{equation}
Another ones are the electromagnetic invariant $F_{\mu\nu}F^{\mu\nu}$, the Ricci-tensor quadratic invariant $R_{\mu\nu}R^{\mu\nu}$, the Kretschmann scalar $K$ and the 2D-horizon Gauss curvature ${^{(2)}\!}R/2$. On the horizon (let us focus on ``the first one" without any loss of generality), all these scalars become extremely simply related and reduce to the value
\begin{equation}
  K\eqH 2\,R_{\mu\nu}R^{\mu\nu}=2\,(F_{\mu\nu}F^{\mu\nu})^2\eqH 2{^{(2)}\!}R^2
   \eqH 8/(M_{1})^4
\end{equation}
which is not affected by the other black hole -- it neither depends on the other-hole mass $M_2$ nor on the separation $2b$. The same observation also applies to all other (higher-power) scalars obtained from the Riemann or/and Ricci tensor, for example
\begin{align*}
  &R^\alpha_\beta R^\beta_\gamma R^\gamma_\delta R^\delta_\alpha
     \eqH \frac{4}{(M_1)^8} \,, \\
  &R^\alpha_\beta R^\gamma_\delta
   {R^{\beta\delta}}_{\kappa\lambda}{R^{\kappa\lambda}}_{\alpha\gamma}
     \eqH \frac{8}{(M_1)^8} \,, \\
  &R^\alpha_\beta R^\gamma_\delta
   {R^{\beta\kappa}}_{\alpha\lambda}{R^{\delta\lambda}}_{\gamma\kappa}
     \eqH \frac{4}{(M_1)^8} \,,
\end{align*}
etc.

When speaking about Gauss curvature, we should also recall mean curvature and the main geometric property of horizons: that they are minimal submanifolds. In the Majumdar--Papapetrou (hence stationary) case, the horizon $N\!=\!0$ is a minimal 2D surface within the 3D slicing $\{t={\rm const}\}$, thus it represents the apparent horizon of all these hypersurfaces and its history forms a trapping horizon, an isolated horizon and an event horizon at the same time. Namely, the horizon is a Killing one since the Killing field $\eta^\mu$ becomes null on it.
The main ``quasi-local" property of the horizon is the vanishing of expansion of the outgoing (geodesic) null normal congruence, in other words, vanishing of the mean curvature of the horizon's $\{t={\rm const}\}$-sections within the $\{t={\rm const}\}$ hypersurfaces. Introducing the ``time" unit normal $n^\mu$, the ``radial" unit normal $r^\mu$ and the outgoing null normal $k^\mu$ to the $\{t={\rm const},N={\rm const}\}$ surfaces,
\[n^\mu\equiv\frac{1}{N}\,\eta^\mu \;, \quad
  r^\mu\equiv\frac{1}{\kappa}\,N^{,\mu} \;, \quad
  k^\mu\equiv\frac{1}{\sqrt{2}}\,(n^\mu+r^\mu) \,,\]
the metric of these surfaces is $\tilde{h}_{\mu\nu}=g_{\mu\nu}+n_\mu n_\nu-r_\mu r_\nu$
and the expansion of $k^\mu$, $\tilde{h}^{\mu\nu}k_{\mu;\nu}$, is given by
\begin{align}
  \sqrt{2}\;\tilde{h}^{\mu\nu}k_{\mu;\nu}
  &= \tilde{h}^{\mu\nu}(n_{\mu;\nu}+r_{\mu;\nu})= \nonumber \\
  &= {n^\mu}_{;\mu}+{r^\mu}_{;\mu}-{n^\mu}_{;\nu}n^\nu r_\mu+{r^\mu}_{;\nu}r^\nu n_\mu= \nonumber \\
  &= {r^\mu}_{;\mu}-\frac{\kappa}{N} \;,
\end{align}
since $n_{\mu;\nu}n^\mu=0$ and $r_{\mu;\nu}r^\mu=0$ due to normalisation,
\[{n^\mu}_{;\mu}=\frac{1}{\sqrt{-g}}\left(\sqrt{-g}\,n^\mu\right)_{,\mu}
                =\frac{1}{\sqrt{-g}}\left(\sqrt{-g}\,n^t\right)_{,t}
                =0 \,,\]
and the acceleration of $n^\mu$ reads ${n^\mu}_{;\nu}n^\nu=\frac{N^{,\mu}}{N}$ while the ``acceleration" of $r^\mu$ is perpendicular to $n^\mu$, ${r^\mu}_{;\nu}r^\nu n_\mu=0$.
On our ``first" horizon of the MP space-time, we have
\begin{equation}
  {r^\mu}_{;\mu}\eqH\frac{1}{M_1}\eqH\frac{\kappa}{N} \;,
\end{equation}
so the expansion of $k^\mu$ is really zero there.\\
{\it Note on other simple horizon embeddings:}
(i) Mean curvature of the horizon as a 3D hypersurface $\{N=0\}$ is given just by ${r^\mu}_{;\mu}\eqH\frac{1}{M_1}\,$.
(ii) Mean curvature of the horizon's $\{t={\rm const}\}$-sections within the $\{N=0\}$ hypersurface is zero; this is ``inherited" from the fact that the mean curvature of the $\{t={\rm const}\}$ hypersurfaces themselves vanishes, ${n^\mu}_{;\mu}=0$.

Another significant submanifolds are the meridional planes $\{t={\rm const},\phi={\rm const}\}$, namely the surfaces orthogonal to both Killing directions. Their Gauss curvature, given by half of the Ricci scalar of the respective 2-metric, is quite complicated, but reduces to zero on the horizons. Their mean curvatures, given by the corresponding 2D divergence of the unit normals $n^\mu$ and $\varphi^\mu\equiv\xi^\mu/\sqrt{g_{\phi\phi}}\,$, are zero {\em everywhere}, because ${n^\mu}_{;\mu}=0$ as well as ${\varphi^\mu}_{;\mu}=0$.

We can also add a simple scalar (counterpart of $\kappa$) given by gradient of the second metric invariant $\sqrt{g_{\phi\phi}}$. Resorting to the spheroidal coordinates, we have $N$ given by (\ref{N,spheroidal}) and
\[g^{rr}=N^2, \quad  g^{\theta\theta}=\frac{N^2}{(r-M_1)^2} \;, \quad
  \sqrt{g_{\phi\phi}}=\frac{r-M_1}{N}\,\sin\theta \,,\]
so one easily computes square of the gradient\footnote
{Note that this quantity must approach unity on the symmetry axis in any axisymmetric space-time in order that the parameter $\phi$ of this symmetry be normalized conventionally.}
\begin{equation}
  \lambda^2\equiv g^{ij}\left(\sqrt{g_{\phi\phi}}\right)_{\!,i}
                        \left(\sqrt{g_{\phi\phi}}\right)_{\!,j} \;.
\end{equation}
On the Majumdar--Papapetrou horizon ($r\!=\!r_{\rm H}\!=\!M_1$), the circumferential radius itself equals $\sqrt{g_{\phi\phi}}\eqH r_{\rm H}\sin\theta$, while its gradient squared equals $\lambda^2\eqH\cos^2\theta$. These values are the same as in the Reissner--Nordstr\"om or Schwarzschild space-time, being again independent of $M_2$ and $b$.

The independence of the horizon's basic geometric characteristics of the other-black-hole parameters indicates that the horizon might not differ at all from the Reissner--Nordtr\"om case of a single black hole. Actually, the horizon area is (\cite{Chandrasekhar}, section 113(c))
\begin{align}
  A_{\rm H}
   &= \lim_{r\rightarrow M_1}\int\limits_0^{2\pi}\int\limits_0^\pi
      \sqrt{g_{\theta\theta}g_{\phi\phi}}\;{\rm d}\theta{\rm d}\phi= \nonumber \\
   &= \lim_{r\rightarrow M_1}\int\limits_0^{2\pi}\int\limits_0^\pi
      \frac{(r-M_1)^2}{N^2}\,\sin\theta\;{\rm d}\theta{\rm d}\phi= \nonumber \\
   &= 2\pi\int\limits_0^\pi\lim_{r\rightarrow M_1}\!\!
      \frac{(r-M_1)^2}{N^2}\,\sin\theta\;{\rm d}\theta = \nonumber \\
   &= 2\pi\lim_{r\rightarrow M_1}\!r\int\limits_0^\pi\sin\theta\;{\rm d}\theta
    = 4\pi(M_1)^2 \,,
\end{align}
the proper distance along any $\{t={\rm const},r={\rm const},\phi={\rm const}\}$ meridian reduces, on the horizon, to
\begin{align}
  l_{\rm H}(\theta)
   &=\lim_{r\rightarrow M_1}\int\limits_0^\theta\sqrt{g_{\theta\theta}}\;{\rm d}\theta
    =\int\limits_0^\theta\lim_{r\rightarrow M_1}\!\!\frac{r-M_1}{N}\;{\rm d}\theta= \nonumber \\
   &=\lim_{r\rightarrow M_1}r\int\limits_0^\theta{\rm d}\theta
    =M_1\theta
\end{align}
and the proper azimuthal circumference along the $\{t={\rm const},r={\rm const},\theta={\rm const}\}$ circle yields there
\begin{align}
  o_{\rm H}(\theta)
   &=\lim_{r\rightarrow M_1}\int\limits_0^{2\pi}\sqrt{g_{\phi\phi}}\;{\rm d}\phi
    =2\pi\lim_{r\rightarrow M_1}\!\!\frac{r-M_1}{N}\;\sin\theta= \nonumber \\
   &=2\pi\lim_{r\rightarrow M_1}\!r\,\sin\theta
    =2\pi M_1\sin\theta \,,
\end{align}
as it should be on a perfect sphere.

\subsection{Further symmetries? Only on horizons}

We started this work with the intention to subject a black hole to the strongest possible influence. We expected that the Majumdar--Papapetrou binary could be the best environment in this respect, but now it appears, on the contrary, that its components are not influenced by each other {\em at all}. This seems rather counter-intuitive, but one must realize that extreme black holes are strange objects, in particular, that a proper distance from an extreme horizon to any point in its exterior (also interior) is {\em infinite}. This means that any ``external" source is effectively at infinite distance from it, and also ``explains" its Meissner-like effect. However, our main aim has been to study the external-source effect on the black-hole {\em interior}. One might tend to expect that in the Majumdar--Papapetrou system the interior will remain unaffected as well, but this is not so, as suggested by uniqueness theorems \cite{ChruscielT-07,GonzalezV-11,ChruscielCH-12}.

We may verify it by checking whether there exists, somewhere (in particular, inside the horizon), some other symmetry besides time and axial symmetry. Were the black-hole interior unaffected, it could be described by the Reissner--Nordstr\"om solution, so two more rotational symmetries would have to exist there. The Killing equation
\begin{equation}
  0=k_{\mu;\nu}+k_{\nu;\mu}=k_{\mu,\nu}+k_{\nu,\mu}-2{\Gamma^\alpha}_{\mu\nu}k_\alpha
\end{equation}
can in the static and axisymmetric case be written out in components as
\begin{align*}
  & k_{t,1}=2\Gamma_{\alpha t1}k^\alpha \,, \quad
    k_{t,2}=2\Gamma_{\alpha t2}k^\alpha \,, \\
  & k_{\phi,1}=2\Gamma_{\alpha\phi 1}k^\alpha \,, \quad
    k_{\phi,2}=2\Gamma_{\alpha\phi 2}k^\alpha \,, \\
  & 0=\Gamma_{\alpha tt}k^\alpha \,, \quad
    0=\Gamma_{\alpha\phi\phi}k^\alpha \,, \\
  & k_{1,1}=\Gamma_{\alpha 11}k^\alpha \,, \;
    k_{2,2}=\Gamma_{\alpha 22}k^\alpha \,, \;
    k_{1,2}+k_{2,1}=2\Gamma_{\alpha 12}k^\alpha \,,
\end{align*}
which simplifies further after substitution for Christoffel symbols,
\begin{align}
  & {k^t}_{,1}=0 \,, \quad
    {k^t}_{,2}=0 \,, \quad
    {k^\phi}_{,1}=0 \,, \quad
    {k^\phi}_{,2}=0 \,, \nonumber \\
  & 0=g_{tt,j}k^j \,, \quad
    0=g_{\phi\phi,j}k^j \,, \label{second-row} \\
  & 2g_{11}{k^1}_{,1}=-g_{11,j}k^j \,, \quad
    2g_{22}{k^2}_{,2}=-g_{22,j}k^j \,, \nonumber \\
  & g_{11}{k^1}_{,2}+g_{22}{k^2}_{,1}=0 \,. \nonumber
\end{align}
Looking for some other symmetries than the time and the axial ones, one focuses on non-trivial solutions of the last two rows (i.e. on solutions with at least one of the components $k^1$, $k^2$ non-zero).

Let us discuss the possibilities in the $x^1\equiv r$, $x^2\equiv\theta$ coordinates.
The first of them is $g_{tt,1}=0$, $g_{tt,2}=0$ which is only possible on $r=M_1$ or at $r=M_1+2b$ and $\theta=\pi$, i.e. on the horizons.
The second possibility is given by $g_{tt,2}=0$, $k^1=0$, $g_{\phi\phi,2}=0$, $g_{11,2}=0$, ${k^2}_{,1}=0$; this cannot be satisfied at all, in particular, $g_{tt,2}$ and $g_{11,2}$ cannot vanish simultaneously.
The third possibility arises when the determinant of the (\ref{second-row}) system is zero,
$g_{tt,2}g_{\phi\phi,1}-g_{tt,1}g_{\phi\phi,2}=0$,
\begin{equation}
  \frac{k^2}{k^1}
  =-\frac{g_{tt,1}}{g_{tt,2}}
  =-\frac{g_{\phi\phi,1}}{g_{\phi\phi,2}} \;.
\end{equation}
Besides the symmetry axis ($\theta=0$ or $\theta=\pi$), the determinant only vanishes on the horizons, however.
Whereas the additional Killing symmetries thus prove possible on the horizons, there are none inside the black holes.
Hence, the Majumdar--Papapetrou horizons are the same whether they are multiple or just one (extreme Reissner--Nordstr\"om), but their interiors differ between these two cases. It will be interesting to check this on the behaviour of the basic invariants.

\subsection{Describing the black-hole interior}

In order to extend the Majumdar--Papapetrou metric (\ref{MP-metric}) below some of the horizons, it is sufficient to reverse, in the lapse $N$, the sign of the respective mass, say $M_1$ (see \cite{HartleH-72} or section 113 of \cite{Chandrasekhar}). When the metric is written in the $(\sigma,\zeta)$ coordinates, (\ref{PM-metric-sigma,zeta}), it automatically covers also the black-hole interiors, it is only necessary to select the coordinate ranges accordingly: below the horizons, $\sigma<b$ and $\zeta<-b$ (below the 1st horizon) or $\zeta>+b$ (below the 2nd horizon) -- see figure \ref{sigma-zeta}; the physical manifold ends at singularities lying on the red-colour hyperbola (there the curvature scalars diverge).
In the spheroidal coordinates ($r$,$\theta$), the central curvature singularity lies at a relevant root of the quartic equation $1/N=0$ -- see (\ref{N,spheroidal}). Clearly for $M_2=0$ or $b\rightarrow\infty$ the singularity radius vanishes and for $b\rightarrow 0$ it approaches the value $M_2$, but a general solution is quite lengthy, so we will only give the singularity location at $\theta=\pi$ (direction toward the second black hole) and $\theta=0$ (antipodal direction):
\begin{align*}
  2r_{\rm sing}&(\theta\!=\!\pi)= \\
               & =M_1+M_2+2b-\sqrt{(M_1+M_2+2b)^2-4M_1 M_2} \,, \\
  2r_{\rm sing}&(\theta\!=\!0)= \\
               &=M_1-M_2-2b+\sqrt{(M_1-M_2-2b)^2+4M_1 M_2} \,.
\end{align*}
As followed in the sense of growing second-hole influence, these values start from the Reissner--Nordstr\"om origin $r_{\rm sing}=0$ in the $M_2\rightarrow 0$ or $b\rightarrow\infty$ limit, and both increase with increasing $M_2$ and/or decreasing $b$, reaching $r_{\rm sing}\rightarrow 2M_1$ for $M_2\rightarrow\infty$.

It is possible to illustrate the deviation of the black-hole interior from spherical symmetry explicitely, on the behaviour of some suitable invariant quantities. It suffices to show, in particular, that an invariant behaves differently along the $\theta=\pi$ and $\theta=0$ parts of the symmetry axis (which means along the direction from the singularity {\em toward} the other black hole and {\em away} from it, respectively). The dependence on $r$ itself is {\em not} conclusive, of course, but one can take some invariant which has local extremes (somewhere) on both the inner parts of the axis, compute the {\em values} at these extremes and compare them. The ratio $\kappa^2/N^3$ is an example of such a quantity. It diverges to $-\infty$ both at the singularity and at the horizon and has a local maximum in between. Choosing $M_1\!=\!1$, $M_2\!=\!1$ and $b\!=\!2$, for instance, the value of the maximum on $\theta\!=\!\pi$ is $(-11.682)$, while the value of the maximum on $\theta\!=\!0$ is $(-12.038)$. One would prefer to integrate the invariants along the two interior counter-segments of the axis (and compare the results), but this typically yields divergence due to the infinite proper distance to the (extreme) horizon. The study of particle motion (namely radial motion along the axis) also does not help, since photons spend infinite Killing time to reach/leave the horizon and time-like particles cannot reach the singularity at all, like in the Reissner--Nordstr\"om case (e.g. \cite{Chandrasekhar}, section 40).

The absence of spherical symmetry below the horizon can also be proved on mutual independence of the invariants, namely by showing that their contours do not coincide. Choosing the simplest two of them, $N$ and $\kappa$ (which also determine $R_{\mu\nu}R^{\mu\nu}$), it is sufficient to show, for example, that normals to $N^{-2}\!=\!{\rm const}$ and to $\kappa^2\!\equiv\!N^{,\iota}N_{,\iota}\!=\!{\rm const}$ are not parallel, which means to calculate the vector-product bivector
$(N^{-2})_{,[\alpha}(\kappa^2)_{,\beta]}$ or the related scalar
\begin{align}
  &
  g^{\alpha\gamma}g^{\beta\delta}
  (N^{-2})_{,[\alpha}(\kappa^2)_{,\beta]}\,(N^{-2})_{,[\gamma}(\kappa^2)_{,\delta]}=
  \nonumber \\
  &
  =\frac{1}{2}\,g^{\sigma\sigma}g^{\zeta\zeta}
   \left[(N^{-2})_{,\sigma}(\kappa^2)_{,\zeta}-(N^{-2})_{,\zeta}(\kappa^2)_{,\sigma}\right]^2=
  \nonumber \\
  &
  =\frac{(24 M_1 M_2)^2 N^{14}}{2\,(\sigma^2-\zeta^2)^{12}}\,
   (\sigma^2-b^2)(b^2-\zeta^2)\;\times
  \nonumber \\
  &
  \quad \times
   \left[M_1(\sigma+\zeta)(\sigma\zeta+b^2)+M_2(\sigma-\zeta)(\sigma\zeta-b^2)\right]^2.
\end{align}
We have evaluated the expression in the $(\sigma,\zeta)$ coordinates were the lapse has the simplest form (\ref{N,sigma-zeta}). It is clear that the result only vanishes at special locations, not on any whole domain (like everywhere below horizon). One however expects the contours to coincide on the axis and this is really the case, because
\begin{align*}
  \theta=\pi \;\; &\Rightarrow  \;\;
                          {\rm for}\;\,r<M_1+2b: \;\, \sigma=b,\;\zeta=b+M_1-r, \\
             &\quad \quad {\rm for}\;\,r>M_1+2b: \;\, \sigma=r-M_1-b,\;\zeta=-b, \\
  \theta=0   \;\; &\Rightarrow  \;\; \sigma=b+r-M_1,\;\zeta=b
\end{align*}
\[\hspace*{-15mm}
  \Longrightarrow \quad
  \sigma^2-b^2=0 \quad {\rm or} \quad b^2-\zeta^2=0 \;,\]
so the above vector-product square vanishes there.

\begin{figure}
\includegraphics[width=\columnwidth]{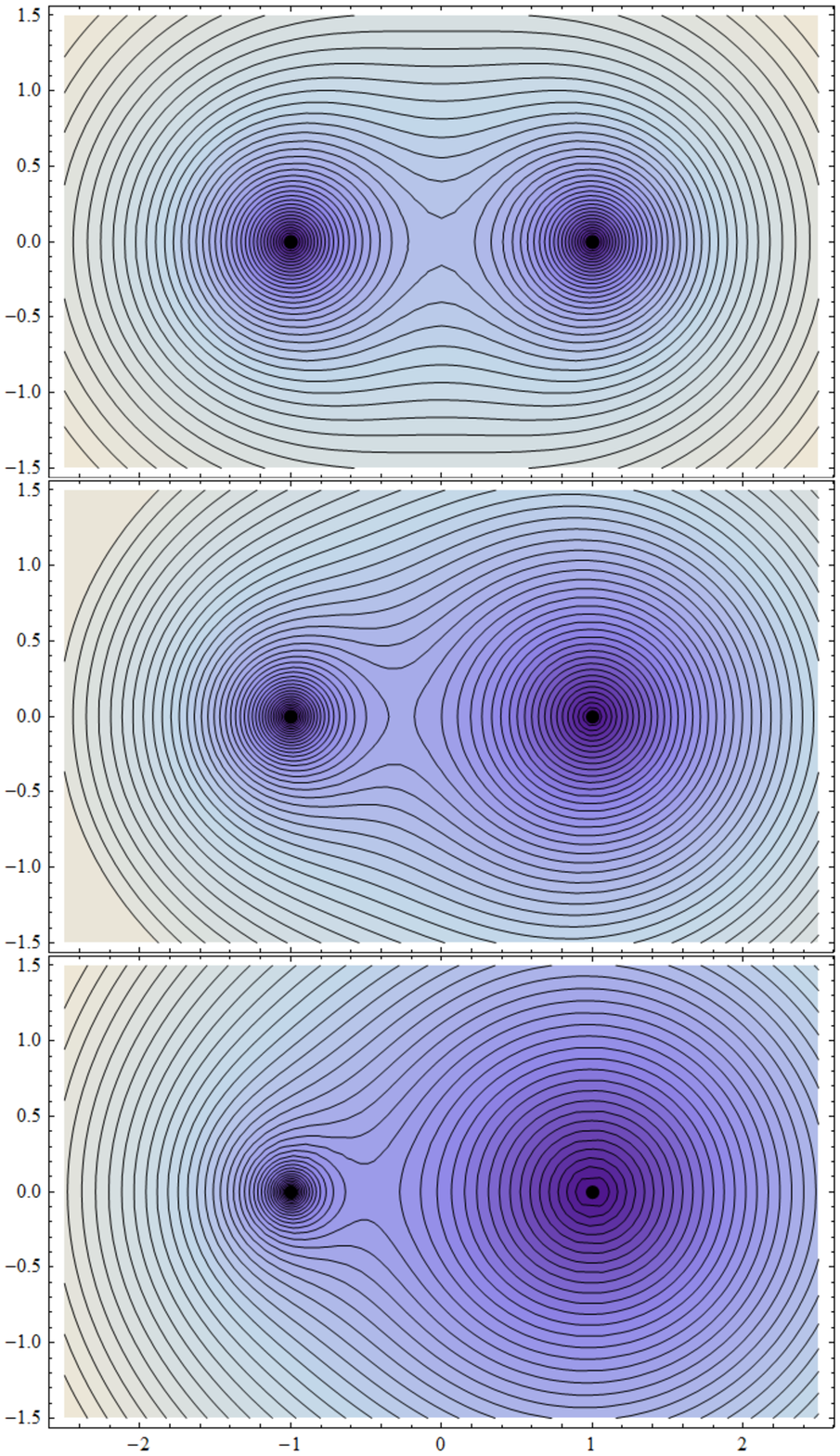}
\caption
{Lapse function $N$ in the outer region of the meridional plane of the Majumdar--Papapetrou space-time with two black holes of masses $M_1$, $M_2$ at coordinate separation $2b=2M_1$. The second-hole (the right-one) mass is $M_2=M_1$ (top) $M_2=3M_1$ (middle) and $M_2=8M_1$ (bottom). Cartesian-type coordinates are used, with horizons represented as points at $x=\mp b$, $y=0$ and axes given in the units of $M_1$.}
\label{MP-lapse-outside}
\end{figure}

\begin{figure}
\includegraphics[width=\columnwidth]{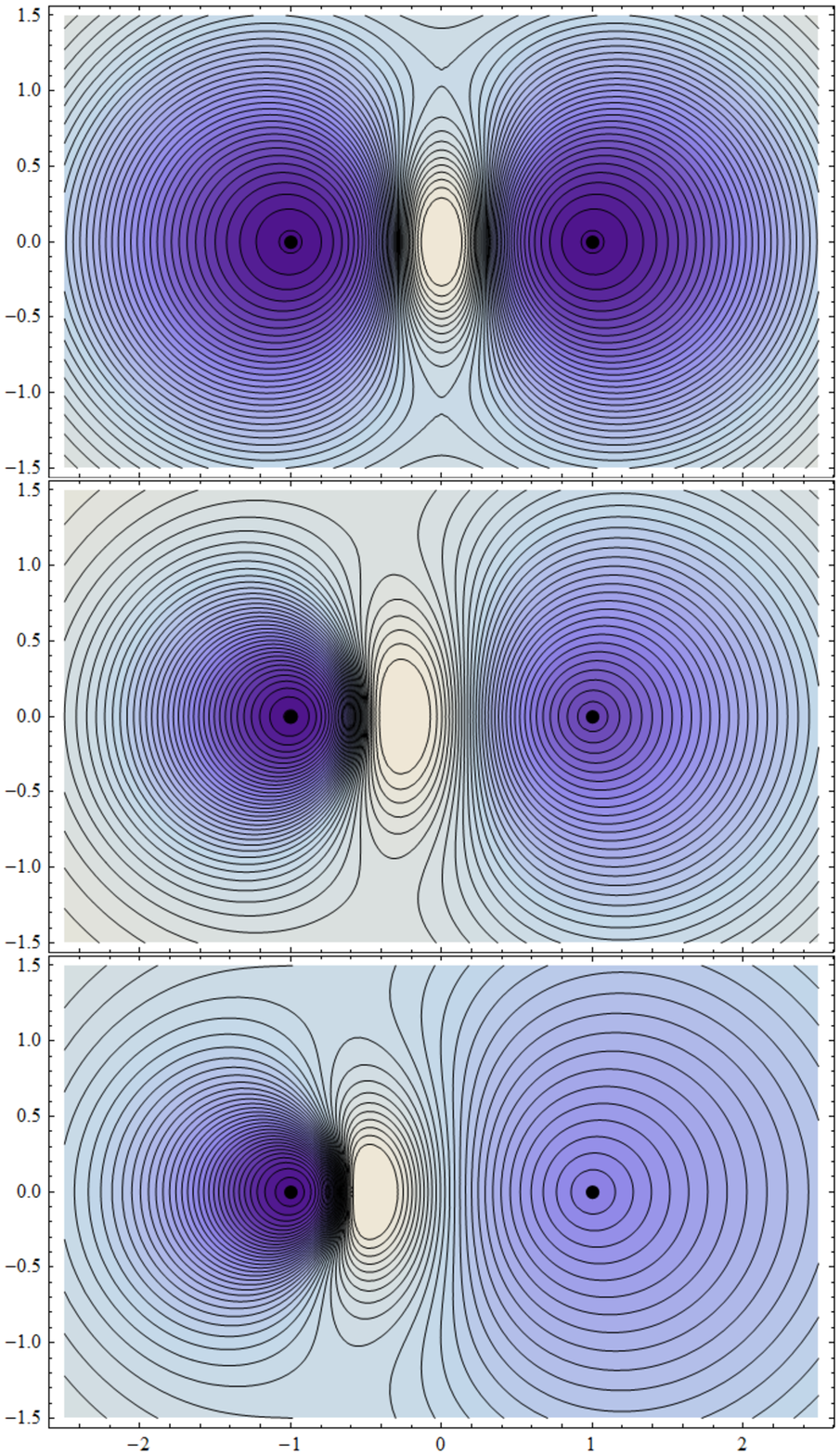}
\caption
{$R_{\mu\nu}R^{\mu\nu}$ scalar in the outer region of the meridional plane of the same binary Majumdar--Papapetrou space-times as in figure \ref{MP-lapse-outside} (mass ratios ``right/left"$\equiv M_2/M_1=1$, $3$ and $8$), plotted in the same way as there.}
\label{MP-Ricci2-outside}
\end{figure}

\begin{figure}
\includegraphics[width=\columnwidth]{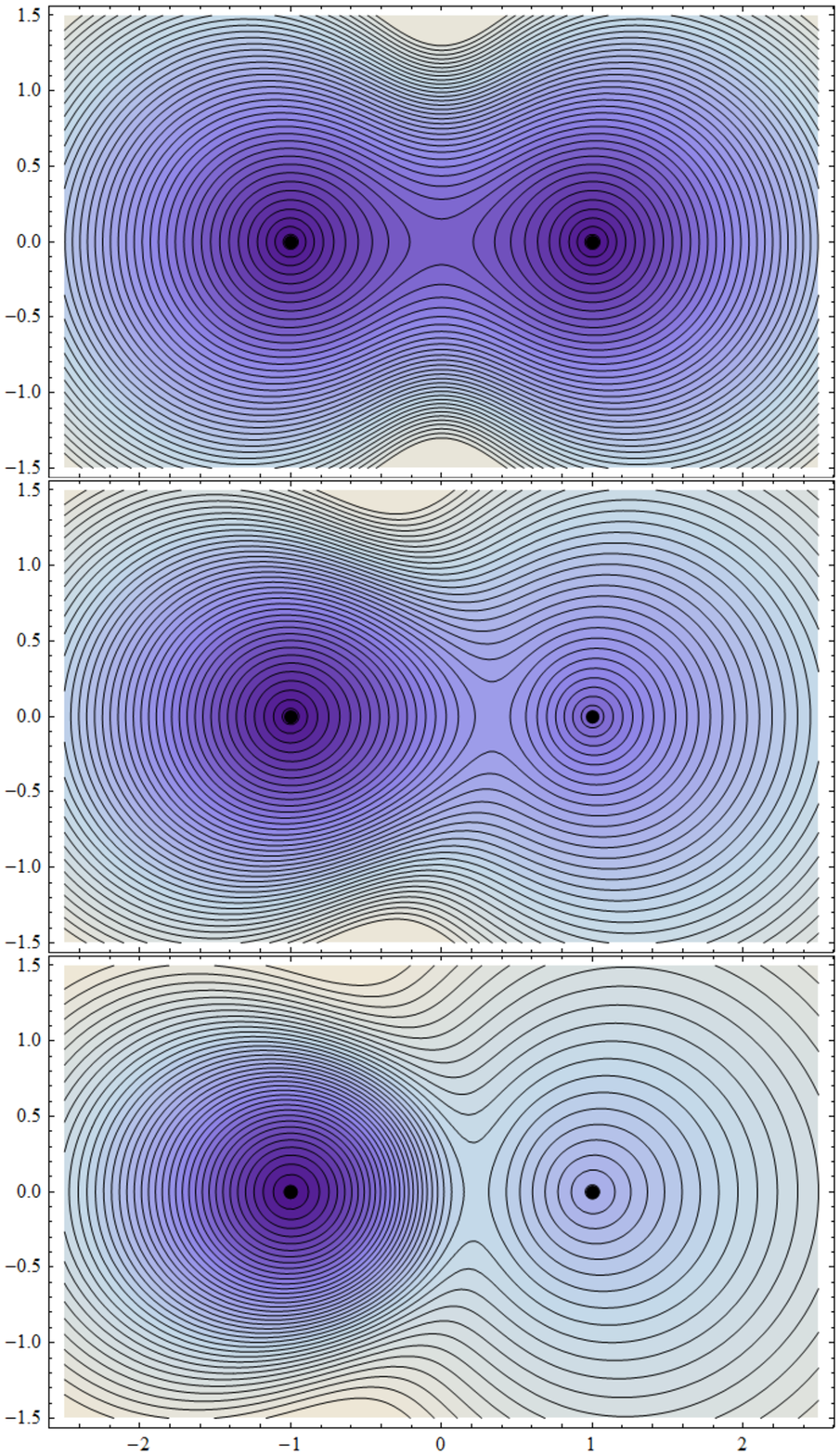}
\caption
{Kretschmann scalar in the outer region of the meridional plane of the same binary Majumdar--Papapetrou space-times as in figures \ref{MP-lapse-outside} and \ref{MP-Ricci2-outside} (mass ratios ``right/left"$\equiv M_2/M_1=1$, $3$ and $8$), plotted in the same way.}
\label{MP-Kretschmann-outside}
\end{figure}

\subsection{Numerical illustrations}

\begin{figure}
\includegraphics[width=\columnwidth]{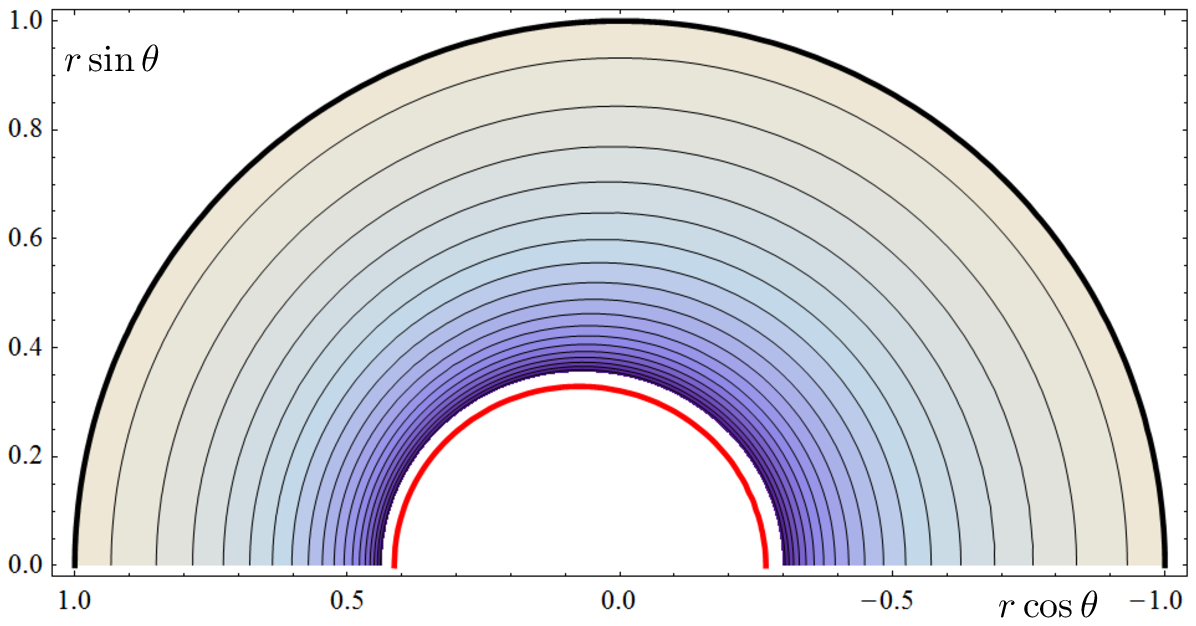}
\caption
{Kretschmann scalar below the ``first" horizon of the same binary Majumdar--Papapetrou space-times as in figures \ref{MP-lapse-outside}--\ref{MP-Kretschmann-outside}, plotted in the $(r,\theta)$ coordinates adapted to the first horizon. The meridional half-section is shown, with the horizontal axis representing the symmetry axis and the second black hole lying to the right of the plot ($\theta=\pi$). The axes are in the units of $M_1$. The outermost curve (the $r=M_1$ circle) is the horizon and the innermost (red) curve is the singularity.}
\label{MP-Kretschmann-inside}
\end{figure}

\begin{figure}
\includegraphics[width=\columnwidth]{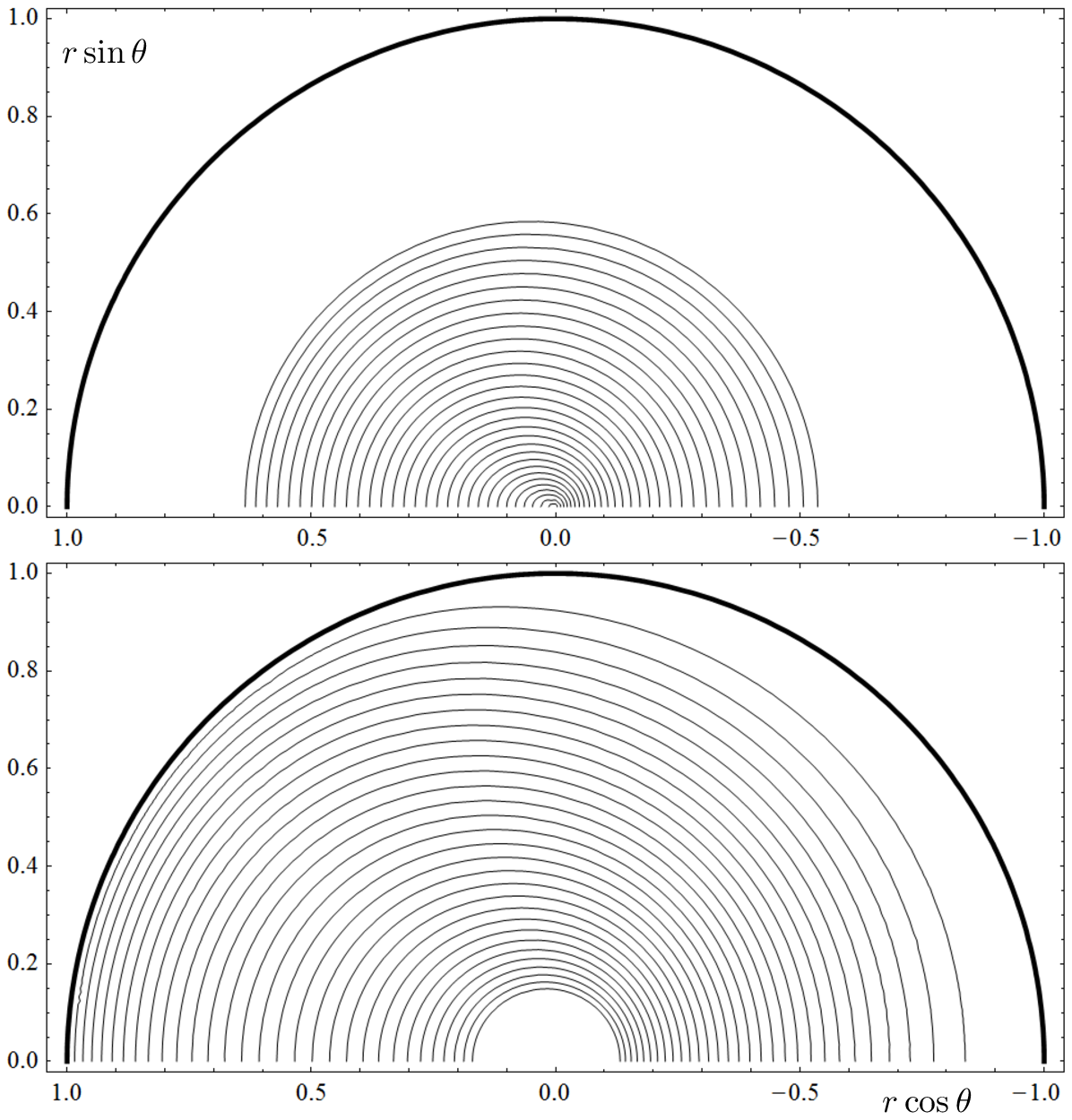}
\caption
{Location of the singularity below the ``first" horizon of the binary Majumdar--Papapetrou space-time, in dependence on the mass ratio $M_2/M_1$ (top) and on separation $2b$ (bottom). Meridional half-sections are shown again, in the $(r,\theta)$ coordinates adapted to the first horizon (represented as the thick $r=M_1$ half-circle), with the horizontal axis coinciding with the symmetry axis and the second black hole lying to the right of the plot ($\theta=\pi$). In brief, the singularity radius grows with increasing $M_2$ and/or decreasing $b$.
Numerical details: going from the innermost curve to the outermost one, the singularity location is shown
(i) {\it top:} for $b=M_1$ and $M_2/M_1=3\left({\rm e}^{-j/10}-{\rm e}^{-3}\right)$, $j=29,28,27,\dots,1,0$
(i.e., $M_2/M_1=0.016$, $0.033$, $0.052$, \dots, $2.565$, $2.851$);
(ii) {\it bottom:} for $M_2=M_1$ and $b/M_1=3\left({\rm e}^{-j/10}-{\rm e}^{-3}\right)$, $j=0,1,2,\dots,28,29$
(i.e., $b/M_1=2.851$, $2.565$, $2.307$, \dots, $0.033$, $0.016$).}
\label{MP-singularity}
\end{figure}

Here the shape of the Majumdar--Papapetrou space will be illustrated on contours of the invariants discussed above: the lapse $N=\sqrt{-g_{tt}}$, the trace of the Ricci-tensor square $R_{\mu\nu}R^{\mu\nu}=(F_{\mu\nu}F^{\mu\nu})^2=4\kappa^4/N^4$ and the Kretschmann invariant $R_{\mu\nu\kappa\lambda}R^{\mu\nu\kappa\lambda}$. We consider three cases of mass ratio in order to see how the pattern changes: a symmetrical binary with $M_2=M_1$, the one with $M_2=3M_1$ and the one with $M_2=8M_1$, where $2b=2M_1$ is kept everywhere as coordinate separation of the holes. The plots are presented in figures \ref{MP-lapse-outside}--\ref{MP-Kretschmann-outside}.
As expected, the less massive black hole produces more ``sudden" curvature, namely the curvature invariants reach higher values at its horizon, but fall off more quickly with distance. Between the holes, there always appears a point where $R_{\mu\nu}R^{\mu\nu}=0$. It is located where $N$ has a saddle, thus where the gravitational attraction of the holes is just in equilibrium; the electric field vanishes at that point, too. One finds easily -- by setting $\sigma=b$ (axis between the horizons) and solving $N_{,\zeta}=0$ -- that this ``central" point lies at
\begin{align*}
  &\zeta=\frac{\sqrt{M_2}-\sqrt{M_1}}{\sqrt{M_2}+\sqrt{M_1}}\;b \quad \Longleftrightarrow \\
  &\Longleftrightarrow \quad
  r=M_1+b-\zeta=M_1+\frac{2b\,\sqrt{M_1}}{\sqrt{M_1}+\sqrt{M_2}} \;,
\end{align*}
similarly as in Newtonian treatment. The zero-field location shifts from the 1st horizon ($r=M_1$) toward the 2nd horizon ($r=M_1+2b$) when $M_1$ increases from zero to values much larger than $M_2$.

However, our main aim has been to see how curvature {\em inside} the horizon responds on the external source. We again plotted contours of the same three invariants as above and have observed that the patterns are pretty similar, so we present just the Kretschmann-scalar ``interior landscape" here, this time in spheroidal coordinates (\ref{spheroidal-coords}) adapted to the first horizon (it is a sphere $r=M_1$ in them) -- see figure \ref{MP-Kretschmann-inside}. Finally, figure \ref{MP-singularity} shows the spheroidal-coordinate location of the singularity in dependence on the other-black-hole mass $M_2$ and on separation $b$. The singularity radius grows with increasing $M_2$ and/or decreasing $b$, though, needless to say, the singularity actually remains {\em point-like} in any case (\cite{Chandrasekhar}, section 113(c)), as seen from the metric (\ref{metric-spheroidal}) which contains $1/N$ in all the spatial elements. The above plots indicate that the divergence of invariants at the singularity is not directional (the iso-surfaces approach the singularity uniformly from all directions).

Let us stress/admit that all the plots are drawn {\em in coordinates}, so they do not represent ``true shapes" of the surfaces, especially one cannot directly compare the pictures obtained for different spaces (different $M_1$, $M_2$ or/and $b$). However, isometric embeddings often look much more ``wild", the more so that extreme horizons are involved which lie at proper radial infinity from both sides.

\section{Concluding remarks}
\label{concluding}

In order to check how an external source affects space curvature generated by a black hole, we have considered a Majumdar--Papapetrou binary black hole and studied the behaviour of the simplest invariants given by the metric and its first and second derivatives. Though ``the other black hole" is a very strong source of gravity, the resulting field is not much deformed within this class of space-times, in the sense that the spatial behaviour of the invariants is not altered very significantly. Even the space-time curvature inside the black hole retains its original shape, in particular, the Kretschmann scalar nowhere turns negative. This is probably connected with the extreme character of its horizons: such horizons are factually cut from all the fields, being characterized by zero surface gravity and shifted to effective infinity. However, the other black hole {\em is} felt inside these horizons -- the interior is not spherically symmetric as for a solitary Reissner--Nordstr\"om hole any longer. 

It thus seems more promising to try to distort a black hole which is far from the extreme state. In such a case, the external source has to be supported somehow in order to allow for a stationary configuration rather than falling onto the hole. Omitting solutions which contain artificial singular ``struts", one can resort to hoop stresses or centrifugal force and turn to discs or rings surrounding the hole. In the simplest approximation, such a configuration can be taken static and axially symmetric, which allows for its exact analytical treatment. Therefore, our plan for the next paper is to consider a Schwarzschild-type black hole with a concentric thin ring. Apart from its theoretical interest stemming from the non-linear superposition, such a system may cover at least some features of space-times of real accreting black holes.

Note finally that the most inhomogeneous field is of course generated by point-like sources. However, these cannot stay in static or stationary equilibrium with the black hole (without supporting struts), unless we return to the Majumdar--Papapetrou type of solutions and endow the point with extremal charge (and the black hole as well).

\begin{acknowledgments}
We thank for support from the grants GACR-14-37086G of the Czech Science Foundation (O.S.), and GAUK-369015 and SVV-260211 of the Charles University (M.B.).
\end{acknowledgments}

\bibliography{deformed-BHs-1.bib}

\end{document}